\documentclass[12pt]{article}
\usepackage{epsfig,graphics,amssymb}
\usepackage{cite}
\usepackage{colordvi}

\begin{document} 
\def\check
{\marginpar{\fbox{check!}} }
\def\nicefrac#1#2{\hbox{${#1\over #2}$}}
\textwidth 10cm
\setlength{\textwidth}{13.7cm}
\setlength{\textheight}{23cm}

\oddsidemargin 1cm
\evensidemargin -0.2cm
\addtolength{\topmargin}{-2.5 cm}

\newcommand{\nn}{\nonumber}
\newcommand{\raw}{\rightarrow}
\newcommand{\be}{\begin{equation}}
\newcommand{\ee}{\end{equation}}
\newcommand{\bea}{\begin{eqnarray}}
\newcommand{\eea}{\end{eqnarray}}
\newcommand{\dl}{\stackrel{\leftarrow}{D}}
\newcommand{\dr}{\stackrel{\rightarrow}{D}}
\newcommand{\dd}{\displaystyle}
\newcommand{\Ln}{{\rm \ln}}
\newcommand{\tb}{\tan \beta}
\newcommand{\fr}{\frac}
\newcommand{\barr}{\begin{array}}
\newcommand{\earr}{\end{array}}
\newcommand{\ra}{\rightarrow}
\newcommand{\mr}{{\stackrel{<}{\sim}}}
\def\bi{\bibitem}
\def\lsim{\:\raisebox{-0.5ex}{$\stackrel{\textstyle<}{\sim}$}\:}
\def\gsim{\:\raisebox{-0.5ex}{$\stackrel{\textstyle>}{\sim}$}\:}
\def\gev{\; \rm  GeV}
\def\eg{ {\it e.g.}}
\def\ie{ {\it i.e.}}
\newcommand{\bm}{\boldmath}
\newcommand{\PP}{{\Red \P}}
\newcommand{\PPP}{{\Blue \P}}
\newcommand{\fmk}[1]{\footnote{{\tt  #1}}}
\newcommand{\fdt}[1]{\footnote{{\it  #1}}}
\newcommand{\fn}[1]{\footnote{{  #1}}}
\newcommand{\Ptt}{{\Red \tt}}
\begin{flushright}
LAPTH-1075/04  \\
IFT-2004-28 \\
hep-ph/0410248 \\
\end{flushright}
\vskip1cm

\begin{center}
{\LARGE {\bf Large 2HDM(II) one-loop corrections in leptonic tau decays }}
\end{center}

\vskip 0.5 cm
\centerline{\large {Maria Krawczyk\thanks{} and David Temes\thanks{}
}}
\begin{center}
\vskip 0.5 cm
{\it 1. Institute of Theoretical Physics, University of Warsaw,
ul. Ho\.za 69 \\ 
 Warsaw, 00-681, Poland}\\
{\it 2. Laboratoire de Physique Th\'eorique LAPTH \\
         Ch. de Bellevue, B.P 110, F-74941, Annecy-le-Vieux, Cedex,
         France.\footnote{UMR 5108 du CNRS, associ\'ee \`a
         l'Universit\'e de Savoie}}
\end{center}
\begin{abstract}
The one-loop contributions to the branching ratios for leptonic 
$\tau$ decays are calculated in the  CP conserving 2HDM(II). 
We found that these one-loop contributions, involving both neutral and charged
Higgs bosons, dominate over the tree-level 
$H^{\pm}$ exchange, the latter 
one being totally negligible for the decay into $e$.
The analysis is focused on large $\tan \beta$ enhanced
contributions to the considered branching ratios.
We derive a simple analytical expression for
the one-loop contribution which holds in this case. 
We show that the leptonic branching ratios of $\tau$ are
complementary to the
Higgsstrahlung processes for $h(H)$ and have a large 
potential on constraining  parameters of the model. In this work we provide 
upper limits on Yukawa couplings for both light $h$ and light $A$
scenarios and we derive new  lower  limit on mass of $M_{H^\pm}$ as a
function of $\tan \beta$, which differs significantly from 
what was considered as standard constraint based on the 
tree-level $H^{\pm}$ exchange only. Interestingly we obtain also an upper limit
on $M_{H^\pm}$. For a SM-like $h$ scenario, with heavy and 
degenerate additional 
Higgs bosons,  one-loop corrections disappear.

\end{abstract}

\newpage



\section{Introduction}

The mechanism of electroweak symmetry breaking is the most important
ingredient in the description of elementary particle physics. 
The Standard Model (SM) incorporates the Higgs mechanism
that breaks the electroweak symmetry spontaneously through a neutral
scalar field with non-zero vacuum expectation value. In the minimal
version of this mechanism one scalar $SU(2)_L$ doublet is required,
providing one physical particle: the Higgs boson. The search of this
particle is one of the main aims of high energy physics and current
searches at LEP exclude SM Higgs bosons with masses below $114.1$ GeV
at $95 \%$ C.L.~\cite{PDG}. 
In this context, valuable information about the Higgs
mass will come from analysis of precise measurements of electroweak
observables. The result of these indirect searches gives an upper
bound on the Higgs
mass $M_{H_{SM}} < 219$ GeV at $95 \%$ C.L.~\cite{PDG}, that is of
great importance for future searches.

Models with Two Higgs Doublets (2HDM) are the minimal extensions of
the SM Higgs sector describing all high energy experimental data and
providing new and rich phenomenology. These models can also be
interpreted as effective theories describing low-energy physics in
models with beyond the SM physics at higher scale. This maybe  the case of
the Minimal Supersymmetric Standard Model (MSSM) with heavy
supersymmetric particles. 
A CP-conserving 2HDM contains $5$ physical Higgs bosons, two neutral
scalars, $h$ and $H$, one pseudoscalar $A$, and two charged Higgs
bosons, $H^\pm$(see e.g.\ \cite{Hunter}). The LEP direct
searches for these Higgs
bosons are more complicated than in the SM due to the number of free
parameters involved and, in particular, the existence in 2HDM of 
 one  very light
Higgs boson can not be excluded~\cite{PDG}. On the other hand,
 LEP data 
excludes, for example, neutral
Higgs bosons with masses below $40$ GeV in the regime of large $\tan\beta$, 
 equal 60 or larger~\cite{exp.yukawa}.

In this context, indirect searches for 2HDM effects in
electroweak observables provide important information about the
masses and mixing angles in the Higgs sector. For example,
concerning the charged Higgs, a lower bound $M_{H^\pm} > 490$ GeV can
be set using indirect effects in $b \to s \gamma$~\cite{gambino}, to
be compared with $M_{H^\pm} > 75.5$ GeV coming from the direct LEP 
searches~\cite{exp.charged}. 
In order to
explore the whole parameter space, global fits using different
electroweak observables $\rho$, $R_b$ and $b \to
s \gamma$~\cite{Maria,Cheung} (and also $(g-2)_\mu$ in~\cite{Cheung}),
have been made, constraining large regions of the parameter space and
therefore giving valuable information for future searches.

In this work, a  complete study of one-loop 2HDM effects in the leptonic 
$\tau$ decays is performed for large $\tan \beta$ and 
arbitrary Higgs spectrum, 
extending the results from~\cite{Rosiek} and~\cite{HollikandSack}. 
 
It will be seen that this radiative effects in the branching ratios for 
$\tau \to e \bar \nu_e \nu_{\tau}$ and $\tau \to \mu \bar \nu_{\mu} \nu_{\tau}$
are
larger that the 2HDM tree-level effects in the relevant regions of
parameter space and experimental data will be used to derive new
constraints for  Higgs masses and mixing angles. 


The paper is organised as follows. The Sect. 2 contains a short 
description of
the 2HDM properties and of   results of the experimental searches on Higgs
bosons. In Sect. 3 the leptonic $\tau$ decay data 
are compared with the SM prediction and the 95 \%CL bounds 
for 2HDM contributions to the branching ratios are derived. In Sect. 4 
the 2HDM contributions  
are parameterised. The one-loop 2HDM effects are computed
in Sect. 5 while their numerical analysis is performed in Sect. 6. Finally,
in Sect. 7 we derived the constrains on the 2HDM parameters coming
from leptonic $\tau$ decay data analysis and our conclusions are
summarised in Sect. 8.

\section{CP conserving 2HDM Model II}
\subsection{General properties}
The  Two-Higgs-Doublet Model is the simplest extension of the 
Standard Model with one extra scalar doublet. It contains   three  
neutral and two charged Higgs bosons. 
Here  we consider a simple CP conserving version with a soft $Z_2$-violation,
assuming  
the Yukawa interactions according to the Model II, 2HDM(II), as in MSSM. 
In this model one of the 
Higgs scalar doublet couples to the up-components of isodoublets while the 
second one to the down-components. In this case there are
7 parameters describing the Higgs Lagrangian: four masses for $h,H,A$ 
and $H^\pm$, 
two mixing angles $\alpha$ and $\beta$ (used in form $\sin(\beta-\alpha)$ and 
$\tan \beta=v_2/v_1$), and the $\nu$-parameter,
related to the soft-$Z_2$ violating mass term in the Lagrangian 
($\nu=\Re m_{12}^2/2 v_1v_2$). 
This $\nu$-parameter describes the Higgs selfcouplings if they are
expressed in terms of masses. We stress that none of
these selfcouplings are involved in this analysis directly.
However our results are sensitive indirectly to the $\nu$ parameter as 
this parameter  governs the decoupling properties of the model.

There is the attractive possibility of having a neutral Higgs boson $h$ 
similar to the SM one, 
 and all  other Higgs bosons 
much heavier. This scenario can be realised in two ways, depending
on the value of  $\nu$-parameter. 
For large $\nu$ the additional Higgs boson masses can be very 
large and almost degenerate, since all of such masses arise from 
one large parameter - $\nu$. It is well known that in such case there is 
{\it decoupling} of these
heavy bosons from known particles, i.e.  effects of
these additional Higgs bosons disappear if their masses 
tend to infinity, eg. in the $\gamma \gamma h$ coupling.  
 At small $\nu$ the  large masses of such additional Higgs bosons arise 
from large  quartic  selfcouplings ($\lambda$) in the Lagrangian.
Since these couplings  are bounded from above by the unitarity 
constraints,  so are the  heavy Higgs-boson masses.
According to these bounds heavy Higgs bosons have to be, typically, 
 lighter than 600 GeV~\cite{unitarity}. 
 Therefore, in
this scenario  the additional Higgs bosons can be heavy enough to
avoid direct observation even at next generation of colliders,
although some relevant effects  can appear in the 
interaction of the lightest Higgs boson (non-decoupling) 
\cite{kanemura,pan,gko,gko-decoup}.

Another interesting scenarios that will be intensively studied in this
work are the ones with mass of $h$ or $A$ below  the SM  
 Higgs boson mass limit, $114$ GeV. In particular, the light $A$ scenario is
especially relevant for the description of $(g-2)_\mu$
data~\cite{muon-exp,muon-tau,muon-2hdm}. These
scenarios are possible without conflict with the existing data  within a 
2HDM(II), since  this model  allows for low production rates  
for very light Higgs particles, as will be discussed below.

The 2HDM (II) model is characterised by the couplings of Higgs bosons to the
fermions and  to the EW gauge bosons. Their ratios  to the 
corresponding couplings  in the SM, $\chi^i_j=g^i_j/g_j^{SM}$,
are presented in Table 1. 
Note that for couplings to the EW gauge bosons $V$,
 $$(\chi^h_V)^2+(\chi^H_V)^2+(\chi^A_V)^2=1,$$ and similarly for 
the couplings to fermions \cite{gko}. 
Note also, that  for each neutral Higgs boson (i) we have
$$(\chi^i_u+\chi^i_d)\chi_V^i=1+\chi^i_u\chi^i_d.$$

\begin{table}[hbt]
\caption{\it  Relative couplings, $\chi^i_j=g^i_j/g_j^{SM}$ in 2HDM (II)}
\begin{center}\hspace{-17mm}
\begin{tabular}{|l|c|c|c|}
\hline
&$h$&$H$&$A$ \\
\hline $\chi_V$ & $\sin(\beta-\alpha)$
& $\cos(\beta-\alpha)$ & $0$  \\
$\chi_u$ & \,\,$
\chi_V^h+\cot\beta\chi_V^H$ &
$
\chi_V^H-\!\cot\beta\chi^h_V$
& $-i\cot\beta$ \\
$\chi_d$ & 
$\chi_V^h-\!\tan\beta\chi_V^H$ &
$\chi_V^H\!+\!\tan\beta\chi^h_V$
& $-i\tan\beta$ \\
\hline 
$\chi_{W^-H^+}$ & $\cos(\beta-\alpha)$
& $\sin(\beta-\alpha)$ & 0 \\
\hline
\end{tabular}
\end{center}
\label{couptab}
\end{table}

Note, that for large $\tan \beta$ the couplings to the charged leptons 
(equal to the couplings to the down-type quarks $\chi_d$), relevant for our analysis,  are enhanced. 

In the last row of  Table~\ref{couptab}   the ${W^{\pm}H^{\mp}\phi^o}$ 
couplings, with $\phi^o=h,H,A$, being of interest of this work, are presented.
Here the ratios of such  
couplings to the SM Higgs boson coupling to gauge boson, $\chi_{W^-H^+}^i= 
g_{W^{\pm}H^{\mp}}^i/g_W^{SM}$, are shown. 

It is important to notice the complementarity between the
$\chi_V^{i}$ on one hand and  $\chi_{W^-H^+}^i$ (and $\chi_d^i$ at large
$\tan\beta$) on the other.

\subsection{Experimental constraints on 2HDM}

The most important constraints on the 2HDM(II) parameter space come
from LEP direct searches for Higgs bosons. Concerning light neutral Higgs
bosons production, there are three main processes within
the energy range covered by LEP, namely, the Higgsstrahlung, $e^+e^-
\to Z^* \to Z h$, the associated production, $e^+e^- \to Z^* \to h
A$, and the Yukawa processes, $e^+e^- \to f\bar f \to f\bar f
h(A)$. The two first processes are highly complementary, due to
their dependence on $(\beta-\alpha)$, 
\begin{eqnarray}
\sigma (e^+e^- \to Z^* \to Z h) &=& \sin^2 (\beta-\alpha)
\sigma_{SM} (e^+e^- \to Z^* \to Z H_{SM}) \nonumber \\
\sigma (e^+e^- \to Z^* \to h A) &=& \cos^2 (\beta-\alpha)
\sigma_{SM} (e^+e^- \to Z^* \to Z H_{SM}) \bar \lambda \\
\sigma (e^+e^- \to Z^* \to f \bar f h ) &=& (\chi_d^{h})^2
\sigma_{SM} (e^+e^- \to Z^* \to f \bar f H_{SM}  )\\
\sigma (e^+e^- \to Z^* \to f \bar f A) &=& (\chi_d^{A})^2
\sigma_{SM} (e^+e^- \to Z^* \to f \bar f  H_{SM} ), 
\end{eqnarray} 

where $\bar \lambda = \lambda_{Ah}^{3/2}/[\lambda_{Zh}^{1/2}(12
  M_Z^2/s + \lambda_{Zh})]$, with
  $\lambda_{ij}=(1-m_i^2/s+m_j^2)^2-4m_i^2m_j^2/s^2$,
being  the two-particle phase-space factor. 

\begin{figure}[h]
\begin{center}
\epsfig{file=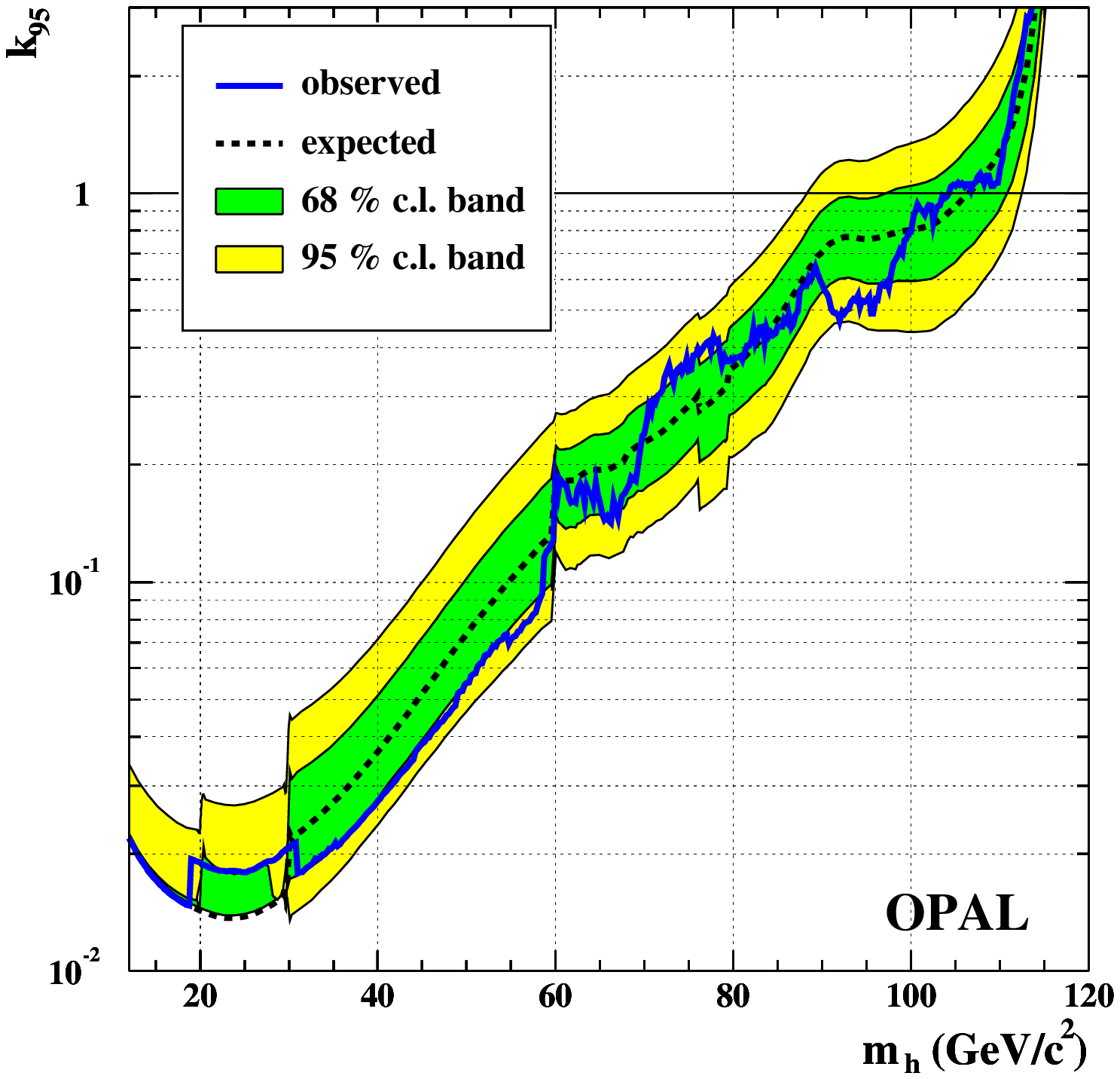,width=7cm}~~
\epsfig{file=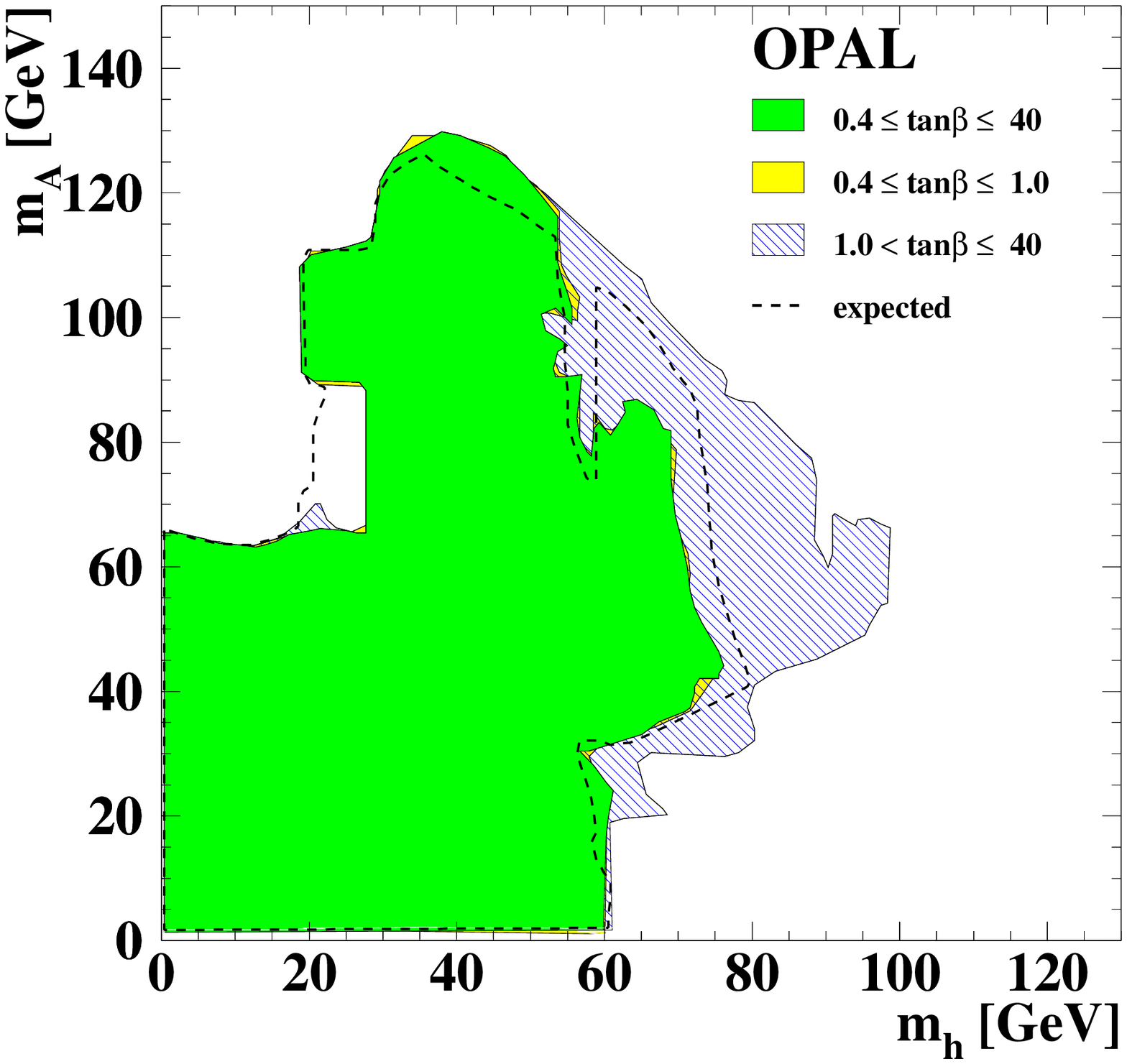,width=7cm}
\caption{{\it Left: Upper limit on $\sin^2(\beta-\alpha)$ as a
    function of $M_{h}$~\cite{exp.Zh}; Right: Excluded
    $(M_{h},M_A)$ regions in the different ranges of $\tan\beta$ for 
    $\alpha = \pm \pi/2,\pm \pi/4, 0$ by OPAL~\cite{exp.hA}.}}
\label{fig.LEP}
\end{center}
\end{figure}

The search for a light $h$ through the Higgsstrahlung process, under an assumption that the light Higgs boson decays into hadronic states, has
been performed in ~\cite{exp.Zh}. The results of this analysis set an
upper limit on the product of the cross section and the corresponding
branching ratio.  It can be translated
into an upper limit on $\sin^2(\beta-\alpha)$ as a function of
$M_{h}$, shown in Fig.~\ref{fig.LEP} 
(left)~\cite{exp.Zh}. Therefore, the results of this analysis are
compatible with a
light $h$ scenario (with mass below 114 GeV) if $\sin^2(\beta-\alpha)$
is small enough.

Also upper limits on the cross section of the associated $hA$ production
process have been derived assuming $100 \%$ decays into hadrons~\cite{exp.hA}.
These results can be translated into forbidden regions in the 2HDM(II)
parameter space. In particular, these results highly constrain 
a scenario with
both $h$ and $A$ light ({\it the light $A\& h$ scenario}). In Fig.~\ref{fig.LEP} the excluded
$(M_{h},M_A)$ regions have been plotted~\cite{exp.hA}. A particular
point is excluded in Fig.~\ref{fig.LEP} (right)  if it is excluded 
for $0.4 \leq \tan\beta \leq 40$
(darker grey region), $0.4 \leq \tan\beta \leq 1$ (lighter grey region), 
and $1 \leq \tan\beta \leq 40$ (hatched region) for all values of $\alpha = \pm
\pi/2,\pm \pi/4, 0$. It is noticeable that a scenario with light $h$
( $A$) is not excluded if $M_A$ ( $M_{h}$) is large enough. In
particular, if $\sin^2(\beta-\alpha)=0$, LEP measurements are
 sensitive to this
associated production if $M_{h} + M_A \leq 130-140$ GeV. 

Finally, the search
for a light Higgs boson has been performed through the analysis of
Yukawa processes assuming that  Higgs boson decays into $\tau$, 
if $2m_{\tau} <
M_{h}, M_A < 2m_b$, or into $b$-quarks, if $M_{h}, M_A > 2m_b$
~\cite{exp.yukawa}. One of the results of this analysis is that 
$M_{h,\, A} \leq 40$ GeV are excluded for high $\tan\beta$
($\tan\beta \geq 60$). We will discuss existing constraints
together with new ones coming from our analysis  in Sect.7. 

\begin{figure}[h]
\begin{center}
\epsfig{file=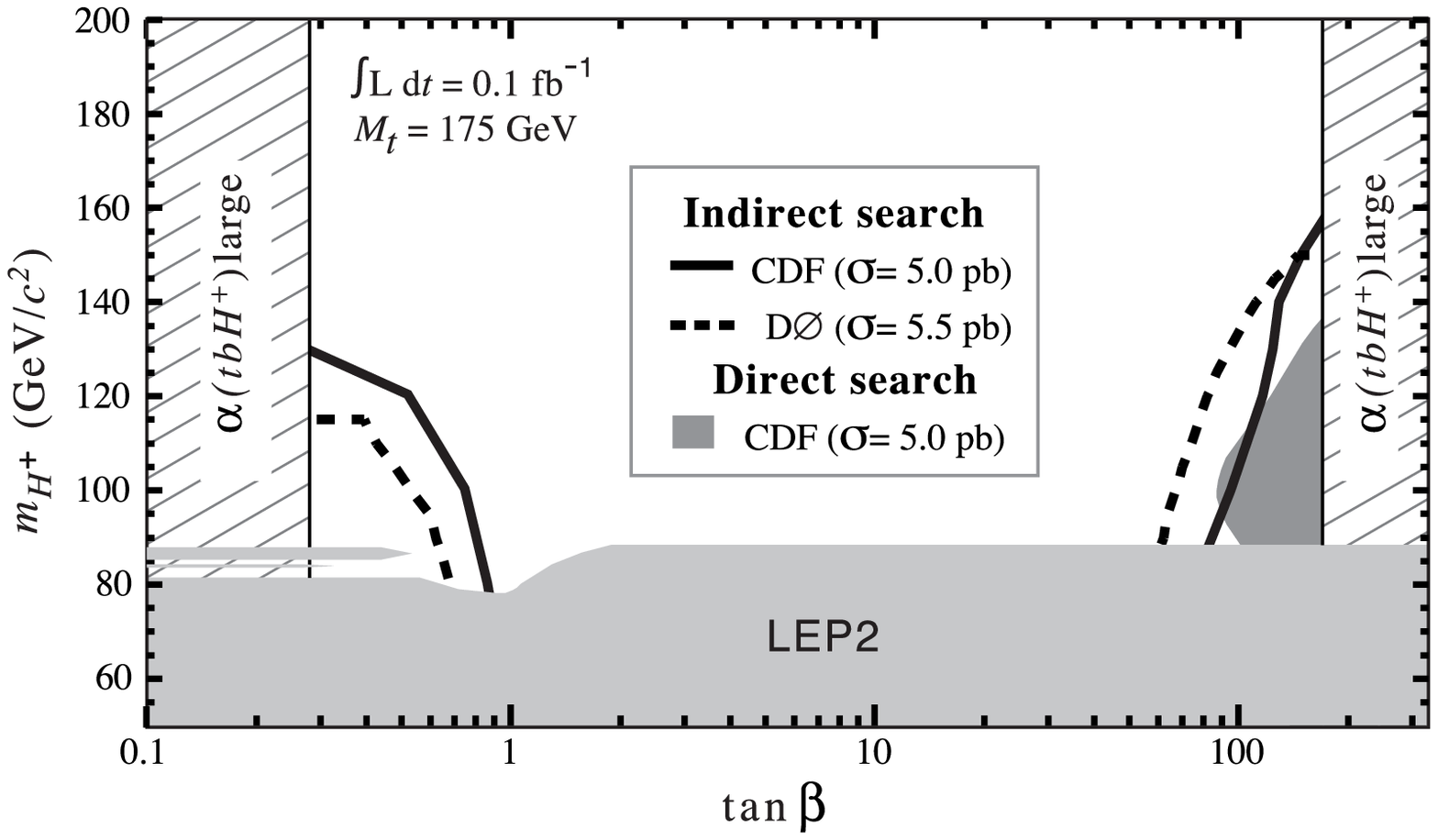,width=12cm}~~
\caption{{\it Constraints on charged Higgs boson mass~\cite{PDG}.}}
\label{fig.TEV}
\end{center}
\end{figure}

Concerning the charged Higgs boson, direct searches at LEP through the
process $e^+e^- \to H^+ H^-$ have been performed assuming $BR(H^- \to
q \bar q) + BR(H^- \to \tau \nu_{\tau}) = 1$. The lower bound 
$M_{H^{\pm}} \geq 75.5$ GeV at $95 \%$
CL~\cite{exp.charged} was obtained. The 
Tevatron data set limits on mass of the charged
Higgs boson as a function on $\tan \beta$, they are presented
 together with LEP results on Fig.~\ref{fig.TEV}.
 
Much stronger constraints on $M_{H^\pm}$ come from the indirect charged Higgs 
boson effects in  $b \to s \gamma$
processes, if interpreted in  2HDM(II).  This leads to a lower
mass limit of $490$ GeV at $95 \%$ for $\tan\beta > 2$\cite{gambino}
\footnote{Recent
  analysis on $B \to X_s \gamma$ predicts larger theoretical errors
  in the SM prediction and therefore a more conservative lower bound
  $M_{H^\pm} \geq 200$ GeV~\cite{Neubert}. }.

In this context, important information on the available 2HDM(II)
parameter space is coming  indirectly   from the low energy
precise measurements. In particular, from the Upsilon decay into $h(A)\gamma$ and  $g-2$ data,
see eg. \cite{muon-2hdm}. 
Also global fits have been performed, combining the results coming
from the different electroweak observables $\rho$, $R_b$ and $b \to
s \gamma$~\cite{Maria,Cheung} (and also $(g-2)_\mu$ in~\cite{Cheung}),
constraining large regions of the parameter space. Here, indirect
constraints of 2HDM(II) will be obtained using leptonic $\tau$ decays
data. The obtained  results will be compared with direct search analysis
coming from LEP and some low energy experiments. 
The implementation of leptonic $\tau$ decay
  data in global fits will be performed elsewhere.

\section{ Leptonic $\tau$ decays: data versus SM predictions}
\label{sect.data}
We consider  the partial decay widths and branching ratios for the two leptonic
decay channels of the $\tau$-lepton, namely
\be
\tau \to e \bar \nu_e \nu_{\tau} \,\,\,\mbox{and}\,\tau \to \mu \bar \nu_{\mu} \nu_{\tau}.\,\,\, 
\ee
We will denote the corresponding quantities using superscript $l$, $l=e$ and 
$\mu$, for example for the branching ratio we use 
$Br^l= Br(\tau \to l \bar \nu_l \nu_{\tau})$.

The '04 world averaged data for the leptonic 
$\tau$ decay modes and $\tau$
lifetime are\cite{PDG}
\be
Br^e|_{exp} = (17.84 \pm 0.06) \%, \hspace{0.5cm}\,\,\,Br^{\mu}|_{exp} = (17.37 \pm 0.06) \% \nonumber 
\ee
\be
\tau_{\tau} = (290.6 \pm 1.1) \times 10^{-15} s.
\ee
Note that the  relative errors of the above measured  quantities are 
of the 0.34-0.38 \%, the biggest being for the lifetime.

The SM prediction for these branching ratios can be defined as the
ratios of the SM predicted
decay widths to the total width as measured in the lifetime
experiments, namely $Br^l|_{SM}
=\Gamma^l|_{SM}/\Gamma^{tot}_{exp}=\Gamma^l|_{SM}
\tau_{\tau}$. Therefore, one can parameterise a possible beyond the SM
contribution by a quantity $\Delta^l$,  defined as 
\begin{equation}
Br^l = Br^l |_{SM} (1 + \Delta^l).
\end{equation}

In the  lowest order of SM the leptonic decay width of the $\tau$  
is due to the tree level  $W$ exchange, see  Fig.~\ref{fig.diagtree} (left). 
Including the W-propagator effect and QED radiative corrections, the 
following results 
for the branching 
ratios in the SM are obtained (see also Sect. 4): 
\begin{equation}
Br^e|_{SM} = (17.80 \pm 0.07)\% ,\,\, \, Br^\mu|_{SM} = (17.32 \pm 0.07)\%.
\end{equation}
Together with the experimental  data this leads  to the following estimations for the possible beyond SM
contributions to the considered branching ratios,
\be
\Delta^e = (0.20 \pm 0.51 ) \% ,\,\,\,\,\,\,
\Delta^{\mu} = (0.26 \pm 0.52) \%.
\ee
Using them we derive the 
 $95\%$ C.L. bounds on $\Delta^l$, for the electron and
muon decay mode, respectively:
\be
(-0.80 \leq \Delta^e \leq 1.21) \% ,\,\,\,\,\,
(-0.76 \leq \Delta^{\mu} \leq 1.27) \%. 
\label{eq.explim}
\ee

One can see that the negative contributions are constrained more strongly
that the positive ones.

\section{Leptonic $\tau$ decays in 2HDM}
In the SM
the leptonic $\tau$ decay,  $\tau \to l \bar \nu_l
\nu_{\tau} $, proceeds at  tree-level via 
the $W^\pm$ exchange.
The formula below describes this contribution in the Fermi
approximation, with leading order corrections to the $W$ propagator,
and dominant QED one-loop contributions,

\begin{eqnarray}
\Gamma^l|_{SM} &= \Gamma^{W^\pm}_{tree}=& \frac{G_{F}^2
  m_{\tau}^5}{192 \pi^3} f(\frac{m_l^2}{m_{\tau}^2}) \left( 1 +
  \frac{3m_{\tau}^2}{5 m_W^2} - 2\frac{m_l^2}{m_W^2} \right) \times
  \nonumber \\
&& \left( 1 + \frac{\alpha(m_{\tau})}{2\pi} (\frac{25}{4} - \pi^2) \right), 
\end{eqnarray}
\begin{equation}
f(x)= 1-8x+8x^3-x^4-12x^2 \ln x.
\label{eq.functionf}
\end{equation}

We will denote the SM contribution in short as $\Gamma_0^l$,  
skipping here and below the  superscript $l$ 
if not necessary.

\begin{figure}[h]
\begin{center}
\epsfig{file=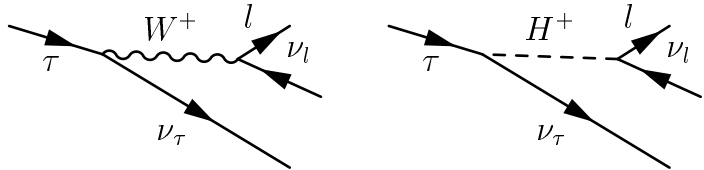,width=14cm}
\caption{{\it Tree-level
    contributions to the $\tau$ leptonic decays. The $W^\pm$ exchange in the 
SM (on the left) and the $H^\pm$ exchange in 2HDM (on the right).}}
\label{fig.diagtree}
\end{center}
\end{figure}


In  2HDM there is, in addition, a tree contribution  due to the 
exchange of the charged Higgs boson, Fig.~\ref{fig.diagtree} (right). This new
contribution is given by
 
\begin{equation}
\Gamma_{tree}^{H^\pm} = \Gamma_0 \left[
\frac{m_{\tau}^2 m_l^2 \tan^4\beta}{4 M_{H^\pm}^4}
- 2 \frac{m_l m_\tau \tan^2\beta}{M_{H^\pm}^2} \frac{m_l}{m_\tau}\kappa  
\left(\frac{m_l^2}{m_{\tau}^2} \right) \right],
\label{eq.treeH}
\end{equation}
where 
\begin{equation}
\kappa(x)= \frac{g(x)}{f(x)},\,\,\,\,  g(x)=1+9x-9x^2-x^3+6x(1+x) \ln(x).
\end{equation}

Note that the second term is coming from the interference with the SM 
amplitude and it
 is much more important than the first one,
that is suppressed by a factor $m_{\tau}^2 \tan^2\beta/8 M_{H^\pm}^2$.
Note that such suppression 
  can  be compensated  by a very large $\tan \beta$ only.

In 2HDM there are  also one-loop contributions 
involving neutral as well as charged  Higgs and Goldstone bosons. 
All these contributions are included in the $G_F$ scheme as follows:

\begin{eqnarray}
\Gamma_{1}^l &=& \Gamma_{0}^l (1+ \delta Z_{L\tau} + \delta Z_{Ll} + 
\delta Z_{L \nu_{\tau}} +
\delta Z_{L\nu_l})  + \Gamma_{loops}^{W^\pm}
\nonumber \\
&&  + \Gamma_{tree}^{H^\pm} + \Gamma_{loops}^{H^\pm}  + \Gamma_{CT}^{H^\pm},
\end{eqnarray}
where the first term corresponds to the SM prediction, 
$Z_{L f} = 1+\delta Z_{L f}$ are the 
renormalisation constants for the 
left component of the fermion $f$ and $\Gamma_{loops}^{W^\pm}$ corresponds to 
the one-loop corrections to the $W^\pm$ exchange tree-level amplitude. The 
$H^\pm$ exchange tree-level contribution and its one-loop and counterterm
corrections are described by $\Gamma_{tree}^{H^\pm}$, 
$\Gamma_{loops}^{H^\pm}$ and $\Gamma_{CT}^{H^\pm}$, respectively.

The tree-level $H^\pm$ contribution is numerically small and 
 the radiative corrections to this amplitude will be neglected here. 
Taking this into account
 we will just consider the tree-level contribution 
eq.(\ref{eq.treeH}), implying that
\begin{equation}
\Gamma_{loops}^{H^\pm} = \Gamma_{CT}^{H^\pm} = 0.
\end{equation}


\section{One-loop 2HDM(II)  corrections}

We evaluate, in the 't Hooft-Feynman's gauge,  the 
one-loop contributions coming from the 2HDM(II) to the quantities $\Delta^l$,
using definitions and conventions for one-loop integrals
of \cite{HollikInt}. We will take into account the fact that the ${H^\pm}$
and $W^{\pm}$ masses are very large compared with the leptonic masses
  and external momenta, and we will neglect masses of muon 
and electron
in the loop calculation. This means that the obtained one-loop
corrections are universal, i.e. they do not depend whether decay 
into $e$ or $\mu$ is  considered, so $\Delta_{oneloop}^{\mu}=\Delta_
{oneloop}^{e}=\Delta_{oneloop}$.
Moreover, we will focus on large $\tan\beta$ enhanced contributions.

\subsection{Renormalisation constants}

\begin{figure}[h]
\begin{center}
\epsfig{file=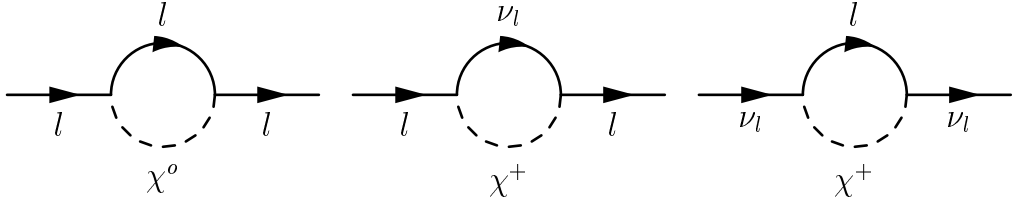,width=14cm}
\caption{{\it Two-point diagrams contributing to the fermion fields
    renormalisation. Here 
$\chi^o = h, H, A, G^o$ and $\chi^+ = H^+,G^+$}.}
\label{fig.diagselfE}
\end{center}
\end{figure}

In order to evaluate the 2HDM contributions to 
the fermion fields 
renormalisation constants, one has to compute the self-energies coming 
from the diagrams shown in Fig.~\ref{fig.diagselfE}.

\vspace{0.5cm}

\paragraph{\bm Charged lepton self-energies.}
There are two kinds of contributions, one involving the exchange of a 
neutral  boson and a second one involving a charged boson. The
latter ones are
 numerically negligible since they are proportional to 
$m_{l}^2/M_{W}^2$ and $m_{l}^2/M_{H^\pm}^2$ (for $\chi^+=G^+$and $ H^+$, 
respectively). 
Therefore  we will 
consider the corrections coming from neutral Higgs and Goldstone bosons only.
Since these corrections are proportional to $m_{l}^2$ 
 we will take into account just the contributions to the 
self-energy of $\tau$. 
We obtain 
\begin{eqnarray}
\delta Z_{L \, e} &=& \delta Z_{L \, \mu} = 0 \nonumber \\
\delta Z_{L \, \tau} &=& \Delta_{\tau}^{h} + \Delta_{\tau}^{H} + 
\Delta_{\tau}^{A} + \Delta_{\tau}^{G^o} 
\nonumber \\
\Delta_{\tau}^{h} &=&  - \frac{G_F m_{\tau}^2}{8\sqrt{2} \pi^2}
\frac{\sin^2\alpha}{\cos^2\beta}  \,\,{\cal B}  (m_{\tau}^2;M_{h}^2,
m_{\tau}^2) \nonumber \\
 \Delta_{\tau}^{H} &=& - \frac{G_F m_{\tau}^2}{8\sqrt{2} \pi^2}
\frac{\cos^2\alpha}{\cos^2\beta}  \,\,{\cal B} (m_{\tau}^2;M_{H}^2,
m_{\tau}^2) 
\nonumber \\
\Delta_{\tau}^{A} &=& - \frac{G_F m_{\tau}^2}{8\sqrt{2} \pi^2}
\tan^2\beta \nonumber  \,\, {\cal B} (m_{\tau}^2;M_{A}^2,
m_{\tau}^2) \nonumber \\
\Delta_{\tau}^{G^o} &=& - \frac{G_F m_{\tau}^2}{8\sqrt{2} \pi^2}
\,\, {\cal B} (m_{\tau}^2;M_{Z}^2,
m_{\tau}^2) \simeq 0,
\end{eqnarray}
where we use the following abbreviation
$${\cal B}()=[B_0 + B_1 + 4 m_{\tau}^2 B_0' + 2 m_{\tau}^2 B_1']().$$

The $G^o$ contribution will be neglected since it is not $\tan^2\beta$ 
enhanced.

\paragraph{\bm Neutrino self-energies.}
In this case only the $H^+$ and $G^+$ contributions are involved and, since 
again these corrections are proportional to the mass of the lepton in
the loop, we will just consider the corrections to the tauonic neutrino field
renormalisation. We obtain 

\begin{eqnarray}
\delta Z_{L \, \nu_e} &=& \delta Z_{L \, \nu_{\mu}} = 0 \nonumber \\
\delta Z_{L \, \nu_{\tau}} &=& \Delta_{\nu_{\tau}}^{H^+} 
+ \Delta_{\nu_{\tau}}^{G^+} 
\nonumber \\
\Delta_{\nu}^{H^+} &=& - \frac{G_F m_{\tau}^2}
{4\sqrt{2}\pi^2} \tan^2\beta [B_0 + B_1] (0;M_{H^\pm}^2,m_{\tau}^2)
\nonumber \\
\Delta_{\nu}^{G^+} &=& \frac{G_F m_{\tau}^2}
{4\sqrt{2}\pi^2} [B_0 + B_1] (0;M_{W}^2,m_{\tau}^2) \simeq 0.
\nonumber \\
\end{eqnarray}

\subsection{One-loop three-point contribution}

\begin{figure}[h]
\begin{center}
\epsfig{file=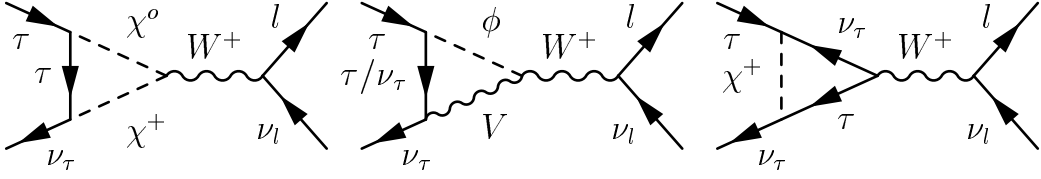,width=14cm}
\caption{{\it Three-point diagrams contributing to the 
$W^{\pm} \tau \nu_{\tau}$
    vertex correction. Similar diagrams exist for the $W^{\pm} l \nu_l$
vertex. $\chi^0=h,H,A,G^o$, $\chi^+=H^+,G^+$ and
    $(V,\phi)=(G^+,Z),(W^+,h),(W^+,h)/(Z,G^+)$}.}
\label{fig.diagtriang}
\end{center}
\end{figure}

The one-loop  three-point diagrams contributing to $\Delta$ 
in the 2HDM(II)  are presented in Fig.~\ref{fig.diagtriang}.
We use here the following notation: $\chi^0=h,H,A,G^o$, 
$\chi^+=H^+,G^+$ and
$(V,\phi)= (G^+,Z),(W^+,h),(W^+,H)/(Z,G^+)$

These $W^{\pm} l \nu_l$ vertex corrections are proportional to the lepton
mass and therefore we will consider only the radiative contributions
to the $W^{\pm} \tau \nu_{\tau}$ vertex. The different
contributions coming from each diagram are as follows.

\vspace{0.5cm}

\paragraph{\bm{$\chi^+-\chi^0-\tau$} {\bm{Loops}}.}
We have computed them (Fig.~\ref{fig.diagtriang} left) 
in the limit of large $M_{H^\pm}$ and $M_W$. That means, 
we have obtained the
complete expressions and keep only  such terms that do not decouple in 
the limit $M_{H^\pm}, \, M_W \gg m_{\tau }$. The resulting expressions are:
\begin{eqnarray}
\Delta_{loops}^{H^+ h} &=& \frac{G_F m_{\tau}^2}{2 \sqrt{2} \pi^2} 
\tan\beta \frac{\sin\alpha}{\cos\beta} \cos(\alpha-\beta) C_{20} 
(m_{\tau}^2,m_{\nu_{\tau}}^2;M_{h}^2, m_{\tau}^2, M_{H^\pm}^2) + \dots 
\nonumber \\
\Delta_{loops}^{H^+ A} &=& \frac{G_F m_{\tau}^2}{2 \sqrt{2} \pi^2} 
\tan^2\beta C_{20} 
(m_{\tau}^2,m_{\nu_{\tau}}^2;M_{A}^2, m_{\tau}^2, M_{H^\pm}^2) + \dots
\nonumber \\
\Delta_{loops}^{H^+ H} &=& - \frac{G_F m_{\tau}^2}{2 \sqrt{2} \pi^2} 
\tan\beta \frac{\cos\alpha}{\cos\beta} \sin(\alpha-\beta) C_{20} 
(m_{\tau}^2,m_{\nu_{\tau}}^2;M_{H}^2, m_{\tau}^2, M_{H^\pm}^2) + \dots
\nonumber \\
\Delta_{loops}^{G^+ h} &=& - \frac{G_F m_{\tau}^2}{2 \sqrt{2} \pi^2} 
\frac{\sin\alpha}{\cos\beta} \sin(\alpha-\beta) C_{20} 
(m_{\tau}^2,m_{\nu_{\tau}}^2;M_{h}^2, m_{\tau}^2, M_{W}^2) + \dots 
\simeq 0
\nonumber \\
\Delta_{loops}^{G^+ H} &=& - \frac{G_F m_{\tau}^2}{2 \sqrt{2} \pi^2} 
\frac{\cos\alpha}{\cos\beta} \cos(\alpha-\beta) C_{20} 
(m_{\tau}^2,m_{\nu_{\tau}}^2;M_{H}^2, m_{\tau}^2, M_{W}^2) + \dots
\simeq 0
\nonumber \\
\Delta_{loops}^{G^+ G^o} &=& - \frac{G_F m_{\tau}^2}{2 \sqrt{2} \pi^2} 
C_{20} (m_{\tau}^2,m_{\nu_{\tau}}^2;M_{Z}^2, m_{\tau}^2, M_{W}^2) + \dots
\simeq 0
\end{eqnarray}

The three last contributions can be neglected in the large $\tan\beta$ limit.

\vspace{0.5cm}

\paragraph{\bm $V-\phi-l$ \bf Loops.}
These contributions (Fig.~\ref{fig.diagtriang}, middle)  
are numerically negligible as do not contain any $\tan^2\beta$ factor.
Therefore, in our work

\begin{equation}
\Delta_{loops}^{V \phi} \simeq 0.
\end{equation}

\vspace{0.5cm}

\paragraph{\bm $\tau-\nu_\tau-\chi^+$ {\bf {Loops}}.}
We have computed these contributions  (Fig.~\ref{fig.diagtriang}, right) 
and checked that they decouple in the limit of very heavy
charged Higgs boson and $W^{\pm}$ boson, as their leading terms in this
limit are proportional to 
$\frac{m_{\tau}^2}{M_{H^\pm}^2}$ or $\frac{m_{\tau}^2}{M_{W}^2}$. Therefore

\begin{equation}
\Delta_{loops}^{\nu_l \chi^+} \simeq 0.
\end{equation}

\vspace{0.5cm}

\subsection{One-loop box diagrams}

The one-loop box diagrams also contribute to the $\tau$ leptonic
decays. All of these diagrams  involve
the exchange of a charged Higgs boson or a
$W^{\pm}$ boson. They can be safely neglected due to the mass dimension of the
$D$ integrals that describe these diagrams, namely
\begin{equation}
D_0 \simeq \frac{1}{M^4} \, , \, D_{\mu} \simeq \frac{1}{M^3}
\, , \, D_{\mu \nu} \simeq \frac{1}{M^2}\, , \, 
D_{\mu \nu \rho} \simeq \frac{1}{M}\, , \, 
D_{\mu \nu \rho \gamma} \simeq {\mathcal O}(M^0).
\end{equation}

Since $M_{H^\pm}$ and $M_W$ are very large as compared to $m_{\tau}$, 
they will drive the mass
dependence of the integrals, so $M=M_{H^\pm}, M_W$. Therefore
only the terms proportional to 
$D_{\mu \nu \rho \gamma}$ do not decouple and give  relevant
contributions. However, in the considered case of $\tau$ decays
there are no
 such contributions. Therefore we can neglect  box diagrams altogether:

\begin{equation}
\Delta_{loops}^{box} \simeq 0
\end{equation}

\subsection{Final expression for one-loop contribution}

\begin{figure}[t]
\begin{center}
\psfig{file=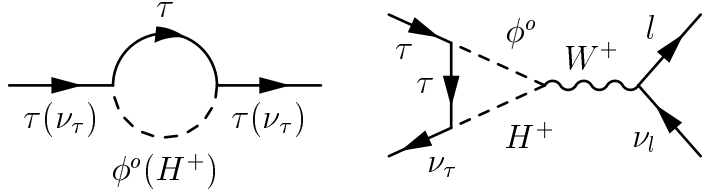,width=14cm}
\caption{{\it Dominant one-loop diagrams contributing to leptonic
    $\tau$ decays in the limit of large
    $\tan\beta$, here $\phi^o = h,\, H,\, A$}.}
\label{fig.diag}
\end{center}
\end{figure}

Taking all this into account,  the dominant diagrams in the limit of large $\tan\beta$ are
  reduced to the ones drawn in Fig.~\ref{fig.diag}. The contributions 
coming from these diagrams are

\begin{eqnarray}
\Delta_{oneloop} &=&  \frac{G_{F} m_{\tau}^2}{8 \sqrt{2} \pi^2}
\tan^2\beta \times \nonumber \\
&&
\left[ 
 - \cos^2(\beta-\alpha)
{\cal B} (m_{\tau}^2;M_{h}^2,
m_{\tau}^2)  \right. \nonumber  \\
&& -  
{\cal B} (m_{\tau}^2;M_{A}^2,
m_{\tau}^2)   \nonumber  \\
&& - \sin^2(\beta-\alpha)
{\cal B} (m_{\tau}^2;M_{H}^2,
m_{\tau}^2)   \nonumber  \\
&& - 2 [B_0 + B_1] (0;M_{H^\pm}^2,m_{\tau}^2)   \nonumber  \\
&& + 4 \cos^2(\beta-\alpha) C_{20} 
(m_{\tau}^2,m_{\nu_{\tau}}^2;M_{h}^2, m_{\tau}^2, M_{H^\pm}^2) 
\nonumber \\
&& + 4 C_{20} 
(m_{\tau}^2,m_{\nu_{\tau}}^2;M_{A}^2, m_{\tau}^2, M_{H^\pm}^2) 
\nonumber \\
&&\left. + 4 \sin^2(\beta-\alpha)  C_{20} 
(m_{\tau}^2,m_{\nu_{\tau}}^2;M_{H}^2, m_{\tau}^2, M_{H^\pm}^2) \right].
\label{eq.total}
\end{eqnarray}

An easy to handle expression can be obtained from 
eq.(\ref{eq.total}) for neutral Higgs masses larger that the $\tau$
mass, $M_{\phi^o} \geq m_{\tau}$. Notice that no
assumption on the Higgs spectrum is made\fn{We generalised here
  the result in \cite{HollikandSack} where 
assumptions were made $M_{H^{\pm}},M_{A} \gg M_{h}$ and
$\alpha=\beta$.}. In this limit, we get \footnote{In agreement
  with~\cite{ChankowskiOld} result derived in the context of the MSSM.}

\begin{eqnarray}
\Delta_{oneloop} \approx \frac{G_{F} m_{\tau}^2}{8 \sqrt{2} \pi^2}
&\tan^2\beta&  \tilde \Delta \nonumber \\
\tilde \Delta=&&
\left[ 
 - \left( \Ln \left( \frac{M_{H^+}^2}{m_{\tau}^2} \right) +
 F(R_{H^{\pm}}) \right) \right. \nonumber \\
&& + \frac{1}{2} \left( \Ln \left(\frac{M_{A}^2}{m_{\tau}^2} \right) + F(R_{A}) \right) \nonumber \\
&& + \frac{1}{2} \cos^2(\beta-\alpha) \left( \Ln\left(\frac{M_{h}^2}{m_{\tau}^2}\right) + F(R_{h}) \right) \nonumber \\
&& \left. + \frac{1}{2} \sin^2(\beta-\alpha) \left( \Ln \left(\frac{M_{H}^2}{m_{\tau}^2}\right) + F(R_{H}) \right) \right],
\label{eq.analyticaltotal}
\end{eqnarray}
where $R_{\phi} \equiv M_{\phi}/M_{H^\pm}$ and
\begin{equation}
F(R) = -1 +2 \,\frac{R^2 \Ln R^2}{1-  R^2}.
\end{equation}

Some useful limits of the $F$ function are:
\begin{equation}
F(R \ll 1) \sim - 1 \, \, \, , \, \, \, F(R=1) = - 3 \, \, \, , \, \, \,
F(R \gg 1) \sim -(1+2\Ln R^2).
\label{flim}
\end{equation}

The above expression
 depends logarithmically on the ratios of the  mass of each
 Higgs boson to the mass of tau lepton. However, 
this may be misleading as in fact there is no dependence on 
mass of the tau lepton in the above expression. 
Indeed, one can see this by looking at other useful form of 
$\tilde \Delta$,\fn{We thank M. Misiak for this suggestion.}
\begin{eqnarray}
\tilde \Delta=
&&3 + \frac{1}{2} \left( G(R_A)+\cos^2(\beta-\alpha)G(R_h)+ \sin^2(\beta-\alpha)
G(R_H) \right), \nonumber \\
\label{eq.analyticaltotal2}
\end{eqnarray}
where 
\begin{equation}
G(R) = \ln R^2 + F(R).
\end{equation}

In the following, to explore the phenomenological
consequences of the large $\tan\beta$ enhanced 2HDM(II) radiative 
corrections to the leptonic  $\tau$ decays,
 we will use both the exact and approximated expressions
(\ref{eq.total}) and (\ref{eq.analyticaltotal}-\ref{eq.analyticaltotal2}),
respectively. 


\subsection{One-loop corrections for some interesting scenarios}

In some phenomenologically interesting scenarios, the expressions 
(\ref{eq.analyticaltotal},\ref{eq.analyticaltotal2}) can be
simplified. 
In the case of light
$h$ and $\sin^2(\beta-\alpha)=0$, $\tilde \Delta$ does not depend on $M_{H}$
and two  limits are worth to be studied, 

\begin{equation}
M_A=M_{H^\pm} \ra \tilde \Delta = \Ln \frac{M_{h}}{M_{H^\pm}} + 1 \, \, \, {\rm {and}} \,\,\, 
M_A \ll M_{H^\pm} \ra \tilde \Delta = \Ln \frac{M_{h}}{M_{H^\pm}} +\Ln \frac{M_A}{M_{H^\pm}} + 2.
\label{twoas}
\end{equation}
Notice, that when $h$ does not couple to gauge bosons and therefore the 
Higgsstrahlung process at LEP is not sensitive to such Higgs boson,
the leptonic tau decays have maximal sensitivity to $h$
as  $\tilde \Delta$  depends logarithmically on its mass, without any
suppression factor.

If $A$ is light and $\sin^2(\beta-\alpha)=1$, one obtains the same
expression for $\tilde \Delta$ that in the previous case with obvious
 replacing $h
\to A$ and $A \to H$. Therefore any analysis with $h$ light and 
$\sin^2(\beta-\alpha)=0$ can be easily translated
 to the case of light
$A$ and $\sin^2(\beta-\alpha)=1$.

The useful expression 
which holds for arbitrary $\sin(\beta-\alpha)$ and degenerate
$H,A,H^\pm$ (with a common mass $M$) is:
\be
\tilde \Delta=\cos^2(\beta-\alpha)\,\,[\ln \frac{M_{h}}{M} + 1]. 
\label{eq.alsba}
\ee

We see that in a SM-like scenario, with light $h$, 
$\sin^2(\beta-\alpha)=1$ and
very heavy degenerate additional Higgs bosons, $\tilde \Delta$ goes to zero,
what signals a  clear decoupling.

\section{Numerical analysis}

In this section we analyse the dependence of the 2HDM(II) one-loop corrections
obtained in the previous sections for the leptonic $\tau$ decays 
on the different Higgs bosons masses and mixing angles. 
First we stress that typically the one-loop contribution dominates 
the 2HDM(II) effects.
They are, for fixed value of large $\tan \beta$ and in the
interesting region of parameter space, five orders
of magnitude larger than the corresponding tree-level $H^{\pm}$ contribution 
to $\Delta^e$,  and one or two
orders of magnitude larger for the $\Delta^{\mu}$. Therefore, although we
will include all contributions in the numerical analysis, the
main features of the 2HDM effects are described by the one-loop
correction in eq.(\ref{eq.analyticaltotal}). In the following only 
results  for muon decay channel will be presented. We stress however
that the obtained  one-loop corrections are the same for  the 
electron and muon channels.

In eq.(\ref{eq.analyticaltotal}) one can distinguish two contributions, one
coming from the 
charged Higgs alone and the other one  
involving also the neutral Higgs bosons. The former is always negative and
it becomes more negative  for a larger charged Higgs mass. 
The latter is typically positive and it grows with the 
neutral Higgs masses. In this way
the total 2HDM(II) one-loop effects, being a sum of two contributions
of the same order and with different signs,
will be large when one of these contributions  dominates. 
Since the modulus of both
corrections grow with the Higgs masses one expects large one-loop effects in 
two cases:
(i) heavy $H^{\pm}$ and light $\phi^o$ (large negative corrections) and 
(ii) light $H^{\pm}$ and heavy $\phi^o$ (large positive
corrections). Taking into
account the lower bound for $M_{H^{\pm}}$ coming from $b \to s \gamma$,
$M_H^\pm$ above $490$ GeV, one
expects to get large radiative effects in case (i) only. 
Note that  in this (i) case the $\tilde \Delta$ (loop contribution) 
is negative, as well as the
tree-level $H^\pm$ exchange, eq.(\ref{eq.treeH}).

We will focus on two scenarios of a special phenomenological interest: 
with a light scalar $h$ and with a light pseudoscalar $A$. 
As all contributions considered here are proportional 
to $\tan^2\beta$, 
they will be plotted for $\tan \beta=1$, to be rescaled by $\tan^2\beta$.


\subsection{Light scalar Higgs boson, $h$ }

First we will consider a scenario with a light scalar boson, $h$, with mass
$M_{h}$ below $114$ GeV, 
and degenerate heavy 
Higgs bosons, with masses $M_{A} = M_{H} = M_{H^+}=M$, above $300$ GeV.
For such  a light Higgs boson $h$, its couplings to gauge bosons
are constrained by LEP data as shown in Fig.~\ref{fig.LEP} (left), 
lying between 0 and $\sin^2(\beta-\alpha)|_{max}$.
Note that for arbitrary $\sin(\beta-\alpha)$ and degenerate
$H,A,H^\pm$ the  equation (\ref{eq.alsba}) holds.

In the light $h$ scenario instead of the degenerated heavy additional Higgs
bosons, one can also consider a spectrum with  SM-like Higgs 
boson $H$ (\ie \,\, with couplings to the gauge bosons as for the SM Higgs, 
namely $\chi^H_V=1$ or $\sin(\beta-\alpha)=0$).
 It is reasonable to assume that such Higgs boson
 has a mass  in the region expected  
for the SM Higgs boson, say $M_{H}= 115$ GeV, 
although as follows from eq. (\ref{eq.analyticaltotal}) 
 nothing depends on this mass. On contrary,   
 one gets here a clear dependence on mass of $h$:

\begin{equation}
\tilde \Delta = \ln \frac{M_{h}}{M} + 1.
\label{eq.applighth}
\end{equation}

\begin{figure}[h]
\hspace{-1.5cm}\epsfig{file=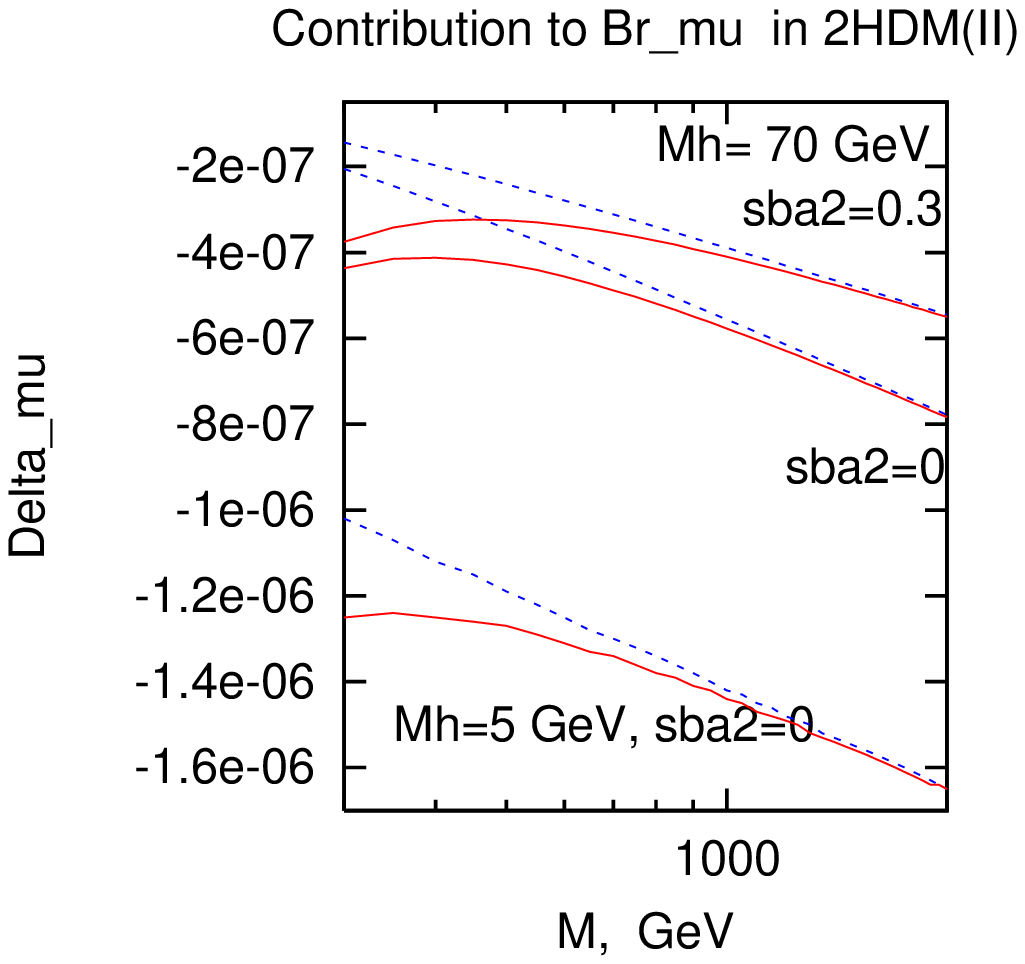,width=8cm}~~
\epsfig{file=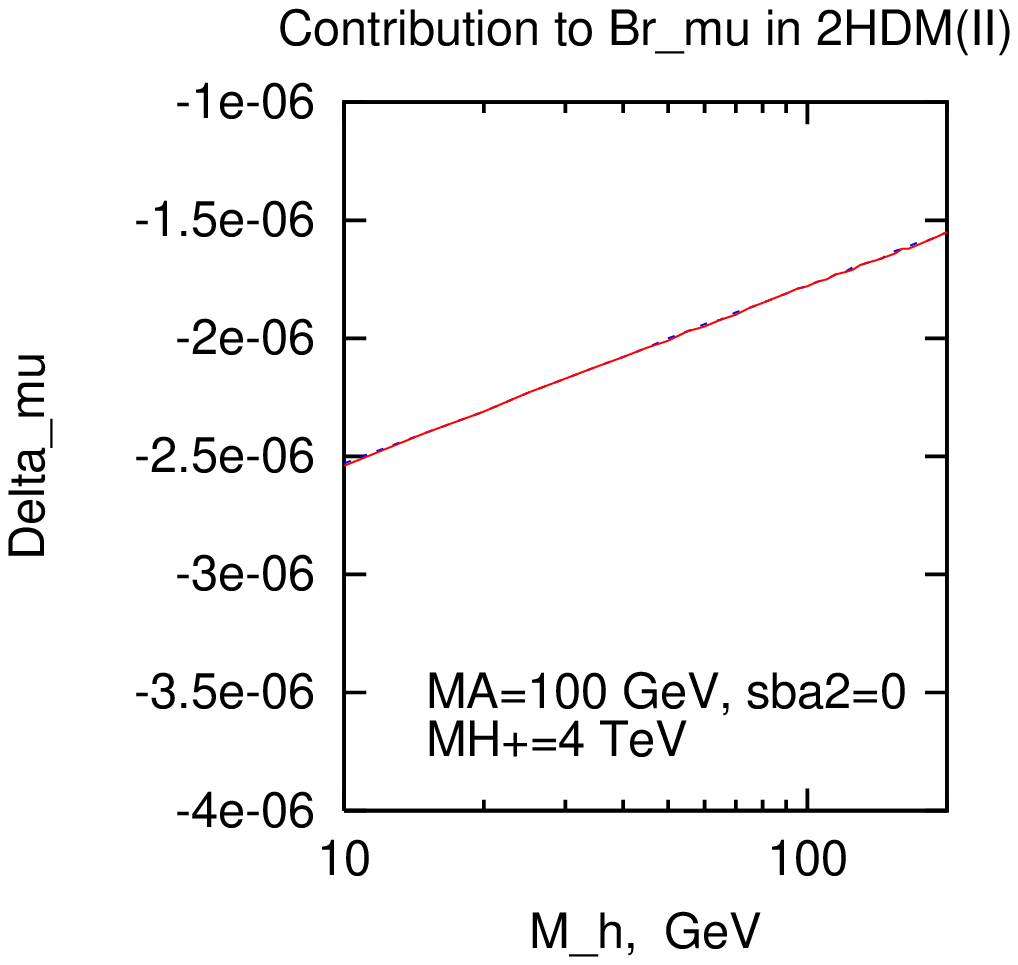,width=8cm}
\caption{\it The total (solid line) and one-loop (dashed line) 
contributions to $\Delta$ for $\tan \beta=1$. Left: Results for $M_{h}=5$
and 70 GeV  are plotted as a 
function of $M=M_A=M_H=M_{H^\pm}$. Results for mass of 70 GeV
for $\sin^2(\beta-\alpha)=0$ and 0.3 are presented by the 
  bottom and upper lines, respectively.
Right: Results  for $M_{A}=100$ GeV, $M_{H^\pm} = 4$ TeV and 
$\sin^2(\beta-\alpha)=0$ ($M_H$ is arbitrary), as a function of $M_h$
are shown.}
\label{fig.lightmh}
\end{figure}

The different contributions to $\Delta \propto \tilde \Delta \tan^2\beta$ 
are plotted in 
Fig.~\ref{fig.lightmh} (left)  for $M_{h}=5$ and 70 GeV, for degenerate 
heavy Higgs bosons.
The total (\ie  \,sum of the tree and one-loop) contributions are plotted using solid lines, while 
the one-loop contributions  using  dashed lines, respectively.
As can be seen, the $H^{\pm}$ tree-level effect is important for low $M$ 
but the one-loop  contribution becomes dominant for $M \geq 500$ GeV. 
In particular, the logarithmic dependence on $M$ coming from the one-loop
corrections is clearly seen. Notice  that curves are plotted for  
$\sin^2(\beta-\alpha)=0$ and $\sin^2(\beta-\alpha)|_{max}$, the maximum 
value allowed by LEP data for a given $M_h$ value. 
For $h$ mass equal to 5 GeV the results for 
different $\sin^2(\beta-\alpha)$, laying  between $0$ and $0.02$,
can not be distinguished.

 The dependence of $\Delta$ on the light Higgs mass can be seen
in Fig.~\ref{fig.lightmh} (left) by
comparing the results obtained for $M_{h} = 5$ and $70$ GeV. This dependence
is explicitly presented in 
Fig.~\ref{fig.lightmh} (right) where the contributions are plotted as
a function of $M_{h}$, for $M_A =
100$ GeV, $M_{H^{\pm}} = 4$ TeV, $\sin(\beta-\alpha)=0$. The 2HDM(II) 
one-loop corrections
decrease logarithmically with  increasing $M_{h}$ as described by 
eq.(\ref{eq.applighth}). So the lighter $h$ the larger the 
one-loop corrections.
One can see that 
$\Delta$ decreases linearly with increasing $\sin^2(\beta-\alpha)$,
in agreement with eq.(\ref{eq.alsba}).

In the case with the SM-like $H$ we have $\sin^2(\beta-\alpha)=0$, then
$\Delta$ becomes insensitive to the value of $M_{H}$, see 
eq.(\ref{eq.analyticaltotal}) and discussion above. Therefore all 
above results
obtained  for the  $\sin^2(\beta-\alpha)=0$ case  hold also here
for the SM-like $H$.

\subsection{Light pseudoscalar, $A$}

In the case where the pseudoscalar, $A$, is light  and the two 
neutral scalars are degenerate $M_{h}=M_{H}$, the
$\tilde \Delta$ does not depend on $\sin^2(\beta-\alpha)$. For
$M_{h}=M_{H}=M_{H^{\pm}}=M$ we get a simple formula
\begin{equation}
\tilde \Delta = \ln \frac{M_{A}}{M} + 1.
\end{equation}
It is  similar to the formula obtained for the discussed above
 case of light $h$ for
 $\sin^2(\beta-\alpha)=0$,
with obvious change $M_h$ to $M_A$. Therefore we will not present the
results corresponding to such light $A$ case.

\begin{figure}[h]
\begin{center}
\epsfig{file=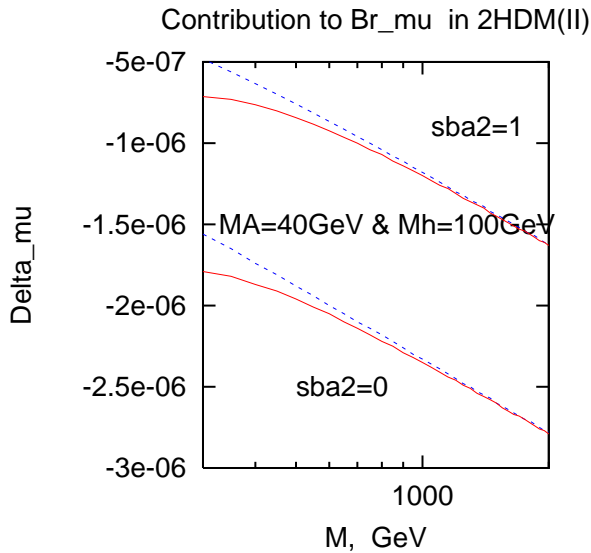,width=8cm}
\caption{\it The total (solid line) and one-loop (dashed line) 
contributions to $\Delta$ for $\tan \beta=1$. Results  for $M_{A}=40$ GeV
and  $M_h=100$ GeV are shown for  $\sin^2(\beta-\alpha)=0$ and $1$,
bottom and upper lines, respectively.}
\label{fig.lightmA}
\end{center}
\end{figure}

There is an interesting  light $A$ scenario where in addition to $A$ 
also $h$ is not very heavy. We call this case
 a light A \& h scenario. 
Here we choose the $h$ mass to be equal to $100$
GeV to  avoid a  direct conflict with LEP data 
presented in Fig.~\ref{fig.LEP} (right). 
In Fig.~\ref{fig.lightmA} the total contribution to $\Delta$
is plotted  as solid lines, while one-loop corrections as dashed 
lines, respectively. Also in this light A \& h scenario
 we see that the one-loop  
effects  dominate for large $M$ scale. The largest  deviation from the SM
prediction occurs for $\sin^2(\beta-\alpha)=0$.

\subsection{Comparison of the exact and approximated results}
Results based on  eq.(\ref{eq.total}) and  
the approximation eq.(\ref{eq.analyticaltotal}) 
have been plotted
together in all the figures,  being clearly  
indistinguishable.  Therefore, the simple approximated  formulae
eqs.(\ref{eq.analyticaltotal},\ref{eq.analyticaltotal2}) can be used
to describe 
the 2HDM(II) one-loop corrections to the $\tau$ leptonic decays
in the whole considered range of parameters.

\section{Constraining 2HDM(II) by the $\tau$ decay data}

In this section we use the leptonic $\tau$ decay data to 
constrain  2HDM(II) parameters.  
The complementarity between LEP processes used for direct searches
of light Higgs bosons and indirect measurements based on the 
  leptonic $\tau$
decays will be exploited to explore the ``pessimistic'', for
the direct searches,  scenarios. In particular the case 
$\sin^2(\beta-\alpha)=0$ will
be studied, since in this scenario the  Higgsstrahlung and VV
fusion processes for  $h$  are suppressed. The 
95 \% CL bounds for
$\Delta$ derived by us in Sect. 3 
 allow to set upper  bounds on $\tan \beta$ (Yukawa couplings) 
for both the light $h$ or $A$ scenarios. We provide also 
exclusion for ($M_h,M_A$) plane for various values of $\tan \beta$ and 
$\sin(\beta-\alpha)$. 

In addition, we  obtain  from $\tau$ decays new lower and 
upper bounds on the charged Higgs boson mass as a function of $\tan \beta$.

\subsection{Constraints  on the Yukawa  couplings of the lightest 
neutral Higgs bosons}

\begin{figure}[ht]
\begin{center}
\epsfig{file=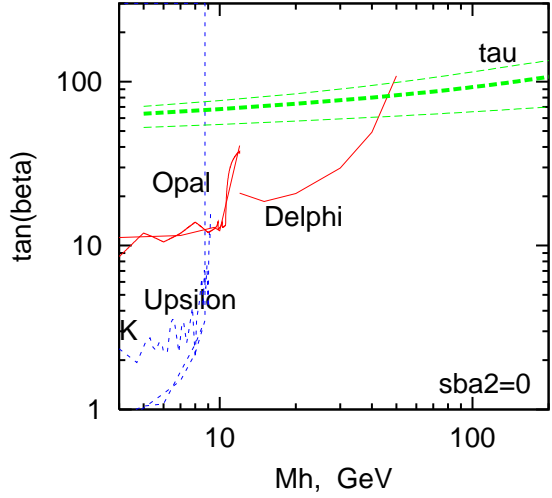,width=8cm}~~
\caption{{\it 95 \%CL upper limits from $\tau$ decay for $\tan \beta$ 
and $\sin(\beta-\alpha)$=0,
 as a function of $M_h$ compared to the existing upper limits from the Yukawa 
processes at LEP (Opal, Delphi) and the Upsilon decay. The two almost 
horizontal lines (in green) corresponds to
$M_A = 100$ GeV, for   $M_{H^{\pm}}=500$ GeV and $4$ TeV, upper and
lower lines, respectively. The results for the degenerate $A$ and $H^+$ with 
mass $4$ TeV are plotted by using thicker line. }}
\label{fig.newmh}
\end{center}
\end{figure}

\begin{figure}[ht]
\hspace{-1.5cm}\epsfig{file=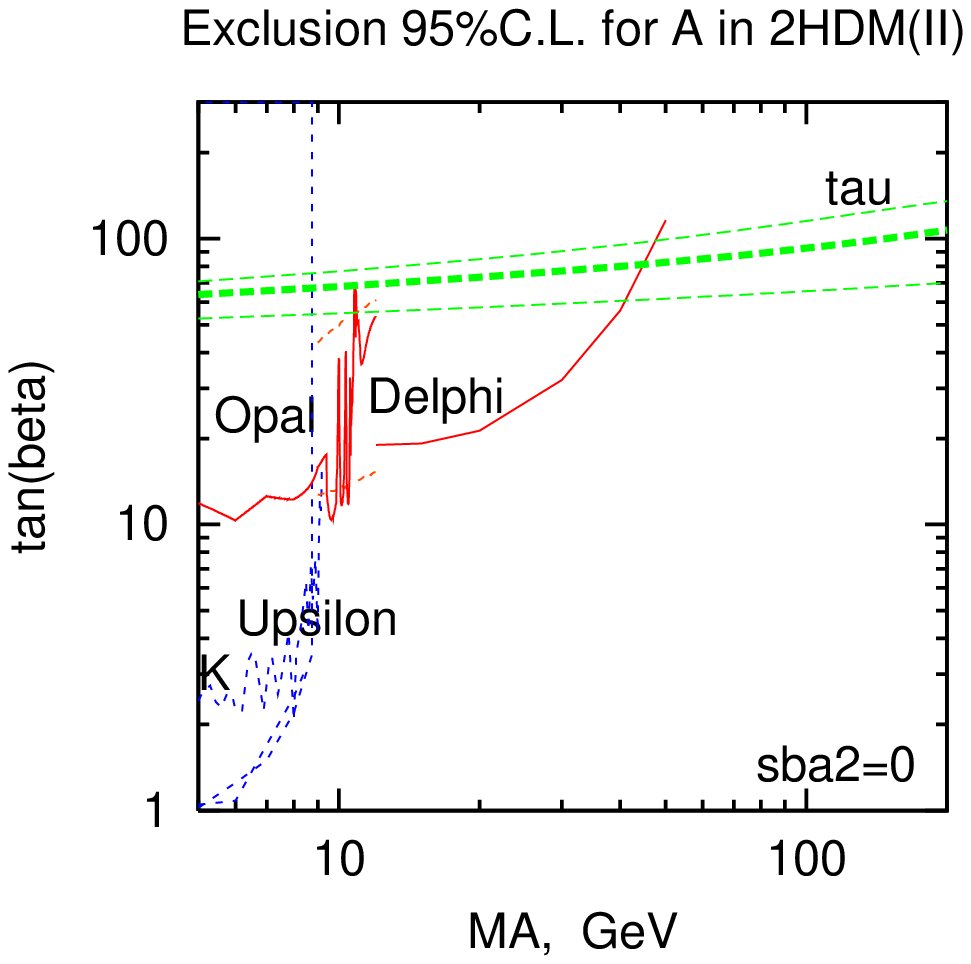,width=8cm}~~
\epsfig{file=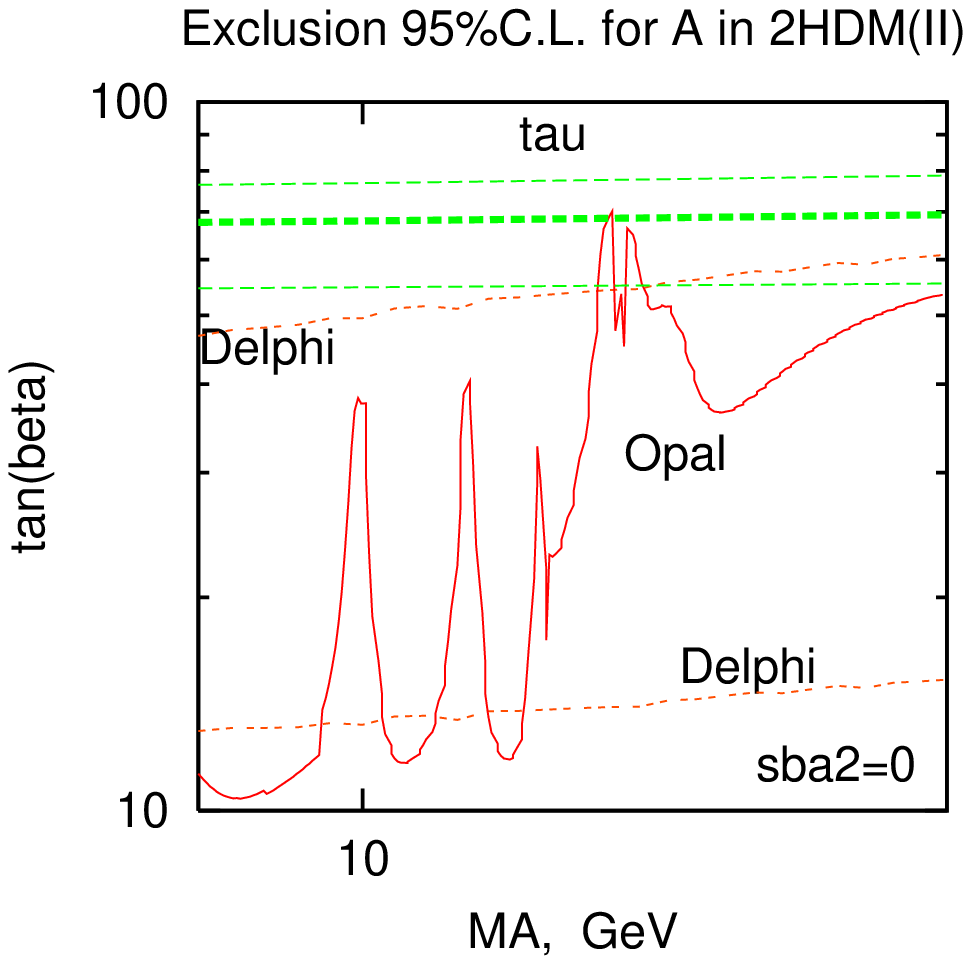,width=8cm}
\caption{{\it 95 \%CL upper limits  from $\tau$ decay for  $\tan \beta$ 
as a function of $M_A$ compared to the existing upper limits from Yukawa 
processes at LEP (OPAL, DELPHI) and Upsilon decay. The two almost horizontal 
lines (in green) corresponds to
$M_h =100$ GeV and $M_{H^\pm}=500$ GeV and $4$ TeV, upper and lower lines,
respectively. The  results for degenerate $h$ and $H^\pm$ with 
mass 4 TeV are plotted in thicker
line. Left: Mass range for $A$ from $5$ to $200$ GeV, Right: Mass range for $A$
from $8$ to $12$ GeV.}}
\label{fig.newmA}
\end{figure}

The upper limits on $\tan \beta$ (Yukawa coupling $\chi_d$)
for light $h$ and light  $A$ scenarios are shown in Fig. \ref{fig.newmh}
and \ref{fig.newmA}, respectively.

In the ``pessimistic'' light $h$ scenario with $\sin^2(\beta-\alpha)=0$, 
the   
leptonic $\tau$ decay data can be exploited to set upper limits 
on the Yukawa couplings as a function of
$M_{h}$. They   can be compared with limits coming from other 
experiments.

In Fig. \ref{fig.newmh} the upper limits on the $\tan \beta$  (Yukawa couplings) for light $h$, assuming  $\sin(\beta-\alpha)=0$, derived from  
the leptonic $\tau$ decay data are presented. One can see that these data 
provide upper 
limits on $\tan\beta$ in region unaccessible by other experiments,
namely for mass above 45 GeV.

As an opposite case to the  light $h$ scenario, one can consider the case with
a light pseudoscalar $A$. If $\sin (\beta-\alpha)=0$, $M_A$
can be low if  $h$ is heavy enough to suppress the associated 
$(h, A)$ production. The Yukawa couplings of $A$ can be  then
constrained just by the Yukawa process with 
$f \bar f A$ final state and the Upsilon decay, $\Upsilon \ra A \gamma$, for a very light 
 Higgs boson $A$. 
Also in this case  the  leptonic $\tau$ decays can be
used to set upper limits on the Yukawa coupling ($\tan \beta$) as a 
function of $M_A$, 
see  Fig.~\ref{fig.newmA}.
The right panel shows the region around mass of $A$ equal 10 GeV.

\begin{figure}[ht]
\begin{center}
\epsfig{file=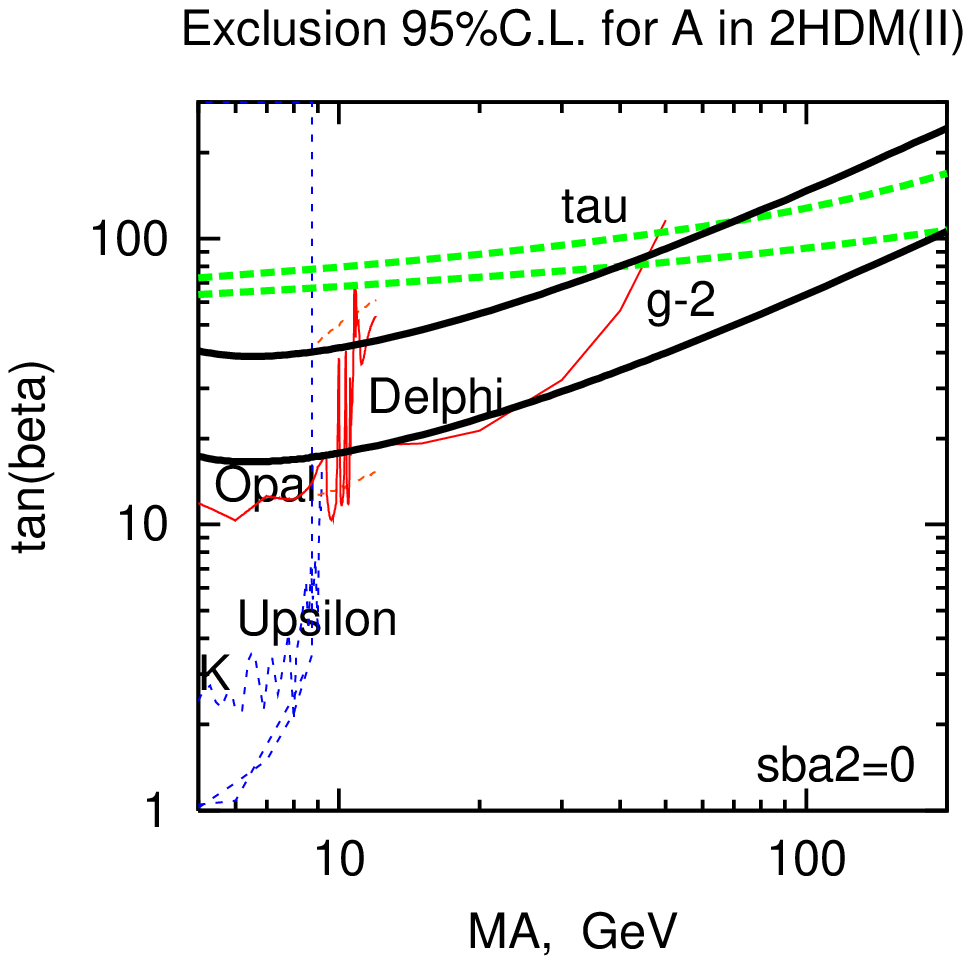,width=10cm}
\caption{{\it Upper limits for $\tan\beta$ from the leptonic $\tau$ decay 
(thick grey lines) and the
    allowed region
from the newest $g-2$ for muon data (thick black lines), in comparison all other existing upper 
limits as a function of $M_A$.
Degenerate masses of $h,H,H^+$ were assumed to be equal to $1$ and $4$ TeV;
the corresponding results from tau decay are given by the  upper and lower 
thick grey lines, respectively.}}
\label{fig.lowmA2}
\end{center}
\end{figure}

Since this scenario can be relevant in explaining $(g-2)_\mu$ data, we plot in
Fig.~\ref{fig.lowmA2} the upper limits for $\tan\beta$ from the leptonic
$\tau$ decay and the allowed region
from the newest $g-2$ for muon data, and for a comparison all other existing 
upper limits for $A$.
Degenerate masses of $h,H,H^+$\fn{ $\sin(\beta-\alpha)$ is 
then arbitrary} were assumed to be equal  to 1 and 4 TeV, the corresponding 
 upper limits are presented by the  upper and lower lines, respectively.

\subsection{Constraints on a light  $A ~\& ~h$ scenario}

A scenario with both $h$ and $A$ light is also of
phenomenological interest. Since $\Delta$ can be large for low
$M_{h}$ and $M_A$, the leptonic $\tau$ decay data can be used to 
constrain   this scenario, in the $(M_{h},M_A)$ parameter space. The
comparison of these constraints with the ones coming from direct
searches will reveal the importance of indirect ones from the  
leptonic $\tau$ decays.

\begin{figure}[h]
\begin{center}
\epsfig{file=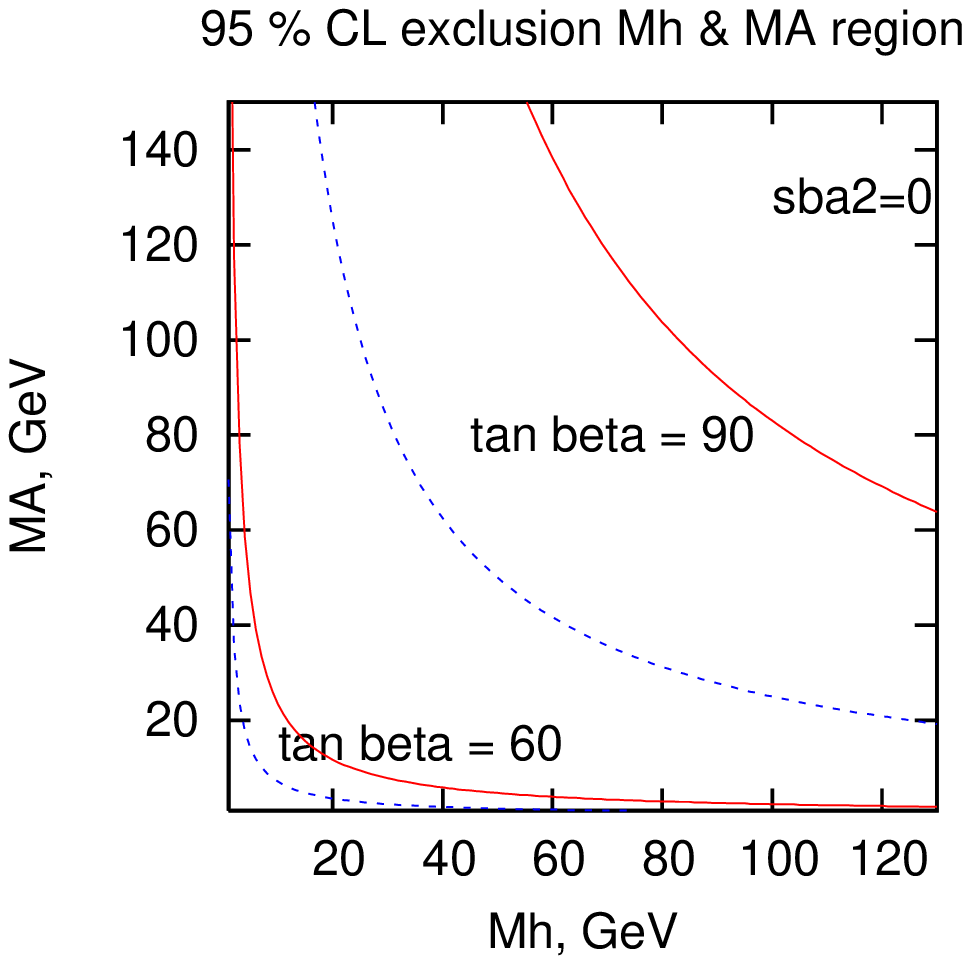,width=8cm}
\caption{\it The excluded regions in the $(M_{h},M_A)$ plane for 
$\sin^2(\beta-\alpha)=0$. The excluded regions lay between axes  and curves
 corresponding to $M_{H^\pm} = 500$ GeV and $1$ TeV, dashed and solid
 lines, and to $\tan\beta=60, \, 90$, down and upper lines, respectively.
}
\label{fig.mhlimit}
\end{center}
\end{figure}

In Fig.~\ref{fig.mhlimit} the constrained regions in the
$(M_{h},M_A)$ plane, laying between axes and the corresponding 
curves, are shown  for $\sin(\beta-\alpha)=0$ and $\tan \beta$ equal 60 and 90.
The excluded regions are symmetric in $M_{h}$ and $M_A$; 
they rule out  the possibility of both $h$ and $A$ being very light. 
These constraints
should  be compared to the constraints shown in Fig.~\ref{fig.LEP} (right). 
For large values of $M_{H^\pm}$ and $\tan\beta$ the 2HDM(II) one-loop effects can
be very large and some of the regions of the parameter space allowed
by direct searches can be excluded indirectly by using
  the leptonic $\tau$ decays.

\subsection{Constraints on the charged Higgs-boson mass}

From the leptonic tau decays one can derive  limits on the mass of 
charged Higgs boson as a function of 
$\tan \beta$. A standard derivation within the 2HDM (II) is based on the 
tree-level $H^+$ contribution for the leptonic tau decay into a muon.
Such derivation can be found in almost all papers,  both 
theoretical  and experimental ones, devoted to this subject (see eg.
 \cite{PDG}). 

First, we apply such a standard method to derive  from  the tree-level contribution (for muon) only the lower mass limit for $H^+$.
By  applying
the obtained lowest value for 
the 95\% CL deviation from the SM prediction (eq.~\ref{eq.explim}) we updated 
the existing lower mass limit. We got  the following limit
\be
M_{H^\pm}\gsim 1.71 \tan \beta \,\,\, \rm{GeV}
\label{eq.treebound}
\ee
with coefficient 1.71  to be compared to the corresponding coefficients
from \cite{dova} and \cite{stahl}, 
equal to 1.86 and 1.4, respectively. Note, that  this is nothing
else, up to the lepton mass ratio, 
what issued as the constraints on the Michel parameter $\eta$ in the 2HDM 
(II), see eg. \cite{PDG,michel}

Next, knowing that the one-loop corrections are typically  more important 
than the tree contribution, we  use them in the derivation of the 
mass limit for $H^+$. We observe that
since $\Delta_{oneloop}$ grows with $M_{H^{\pm}}$, this one-loop correction
allows to put {\it upper
bounds on $M_{H^{\pm}}$} in scenarios with light neutral Higgs
bosons. In particular, for $\sin(\beta-\alpha)=0$, $\tilde \Delta$ goes as 
$\ln(M_{H^{\pm}}/M_{h})+\ln(M_{H^{\pm}}/M_A)$ (eq.~\ref{twoas}) and therefore 
the lighter $h$ and $A$ the stronger upper bounds for $M_{H^{\pm}}$. 

\begin{figure}[t]
\hspace{-1.5cm}\epsfig{file=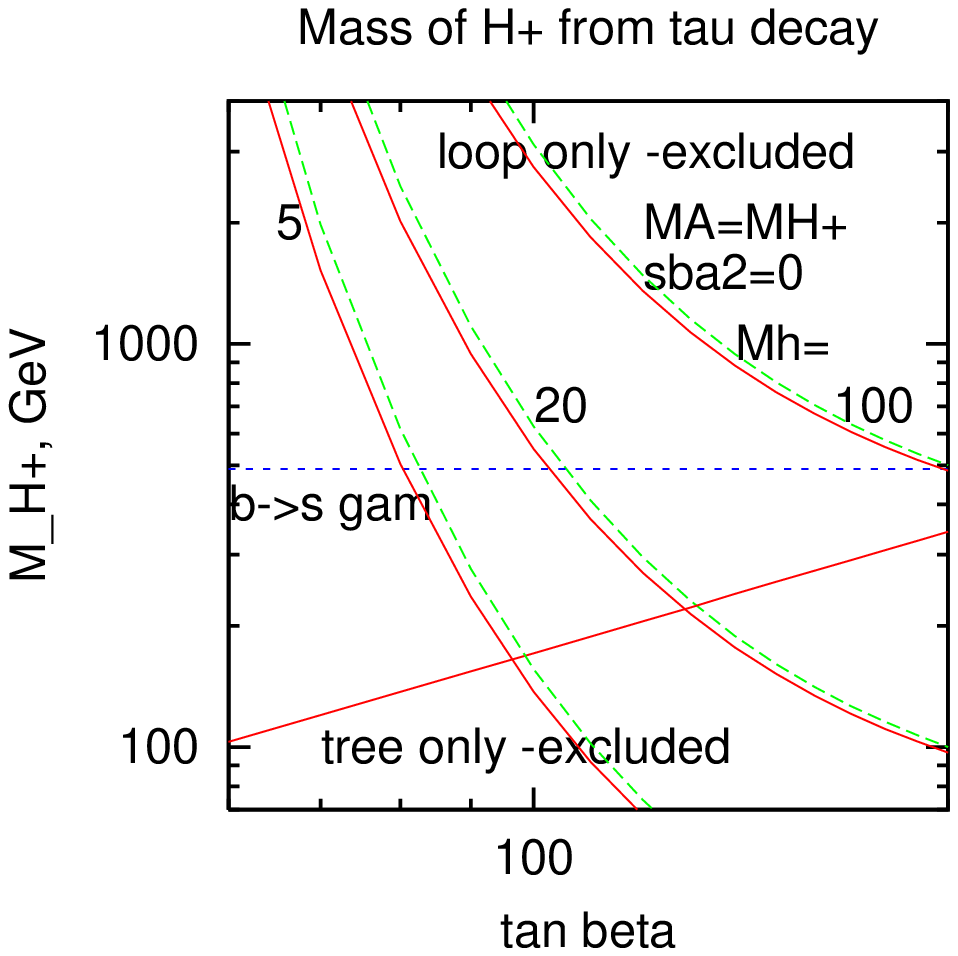,width=8cm}~~ 
\epsfig{file=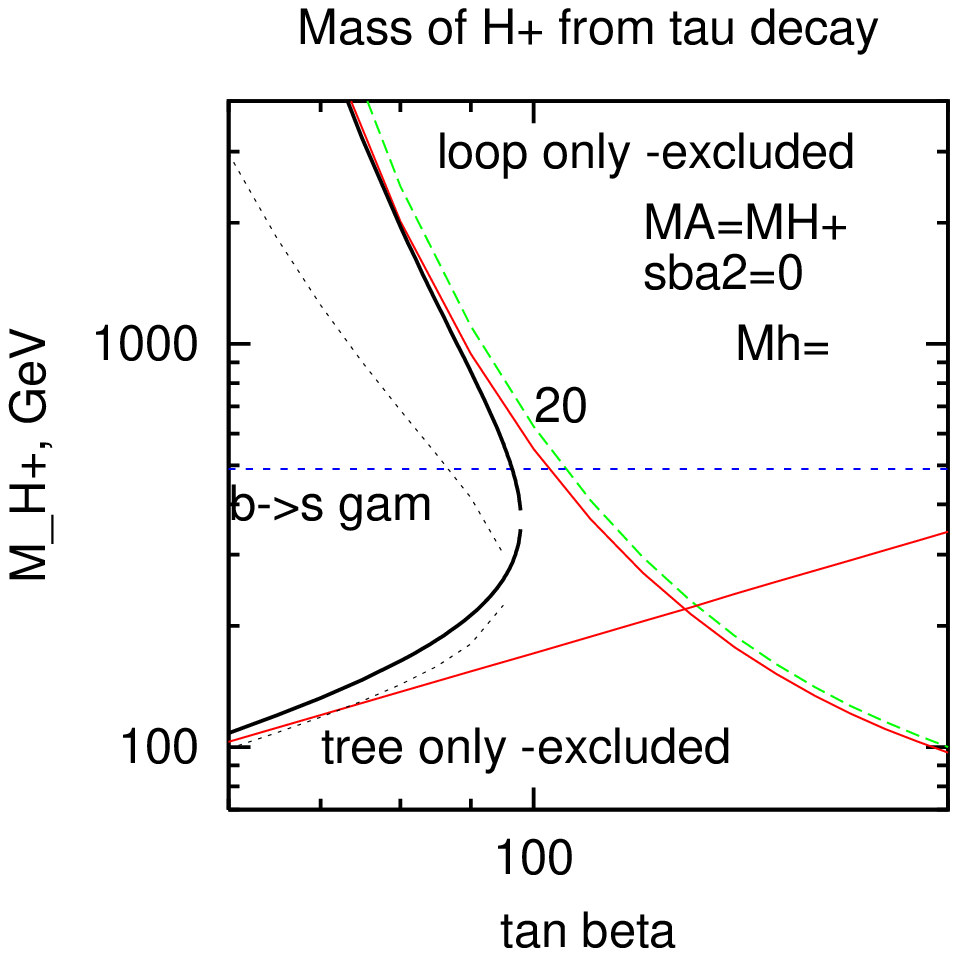,width=8cm}
\caption{\it Limits on charged Higgs boson mass
as a function of $\tan \beta$  obtained from leptonic tau decays. 
The upper limits from  the one-loop contribution 
 (dashed and solid lines correspond to the electron  and muon  channels) 
and lower limits (straight lines) from the tree $H^+$ exchange from muon 
channel are shown. 
Lower limit from $b \to s \gamma$ is also shown.
Left: The upper limits obtained  from the one-loop corrections only
 for $M_{h}=5,20,100$ GeV and 
$\sin^2(\beta-\alpha)=0$, assuming  $M_A=M_H^+$ are presented. 
Right: The same as in Left for one mass $M_h=20$ GeV, in addition
in form of the full constraint from the total (loop plus tree)
contribution, for degenerate masses $M_A=M_{H^\pm}$ 
(thick solid line), and for $M_A=100$ GeV (thin dotted line) 
is presented. }
\label{fig.mhc}
\end{figure}

In Fig.~\ref{fig.mhc} (left) the individual 
lower and upper bounds on $M_{H^{\pm}}$, as obtained from the 
tree-level $H^\pm$ exchange diagram only and from the one-loop contribution only, 
are plotted as a function of $\tan \beta$.
The constraints based on the one-loop contributions are plotted for 
 various masses 
of $h$, equal to  5, 20 and 100 GeV, assuming $\sin(\beta-\alpha)=0$ and a 
degeneracy in masses of $A$ and $H^\pm$.
Upper limits coming from the one-loop corrections are plotted both for the muon and electron
decay channels, the limits from electron one is slightly weaker (dashed (green) lines). 
The relevant lower limits from the tree-level $H^\pm$ contribution is obtained only
from the muon channel.
The lower bound 
coming from $b \to s \gamma$ analysis is also shown for a comparison.

In Fig.~\ref{fig.mhc} (right) we present results as described above 
for one  particular mass of $h$ equal 20 GeV,  
together with a full bound based on  a sum
of tree and one-loop contribution (thick line). We see that a full bound
 gives
both the {\it lower and upper limits} as a function of $\tan \beta$.
In this figure we present also  results  obtained for other  mass of $A$,
equal 100 GeV. 
The thin black dotted line corresponds to the limits obtained then, 
with all other 2HDM parameters as used above to obtain a thick line.

It is clear that the constraints for mass of $H^+$ change drastically 
if the one-loop contributions
are  included in the analysis.  In particular,
 the lower bound is
higher than the tree-level limit, eq.(\ref{eq.treebound}). 
Only for the SM-like $h$ scenario,
 with $\sin(\beta-\alpha)=1$ and all other Higgs boson
mass heavy and degenerate, the tree-level 
 contribution gives a reliable estimation.

Based on results presented in  Fig.~\ref{fig.mhc} the 
restrictions can be set on
$M_{H^{\pm}}$ for large values of $\tan\beta$ ($\tan\beta \geq
60$). In particular, in a scenario with light $h$ and not so heavy
$A$, $M_{H^{\pm}}$ should be lower than $3$ TeV for $\tan\beta =
65$. Although large values of $\tan\beta$
are required, this upper bound to the charged Higgs mass is important
due to the difficulty on setting upper bounds on masses of
undiscovered particles.

\section{Summary and conclusions}

In this work we have computed the 2HDM(II) one-loop corrections to
the leptonic $\tau$ decays. As a main  result we have obtained that
these one-loop effects are larger than the corresponding tree-level $H^\pm$
contribution in the relevant regions of the parameter space. Our
analysis has been focused on the $\tan\beta$ enhanced contributions
and an easy-to-handle formula has been obtained describing these
one-loop effects in the approximation of the  Higgs boson masses larger than the
$\tau$ mass. This formula allows us to study all the 2HDM parameter
space in a transparent  way. 

After the numerical analysis of the
corrections, the constraints on the 2HDM(II) parameters from the
leptonic $\tau$ decay data have been obtained in different
scenarios. In particular the ``pessimistic'' scenarios for the direct
searches of light $h$  Higgs bosons at LEP have been intensively
analysed. From this analysis we have obtained  upper limits on the
Yukawa couplings for both light $h$ and light $A$ scenarios,
constraining also the light  $A \& h$ scenario. We have updated
the existing in literature lower limits on $M_{H^\pm}$, 
different from the ones coming from
tree-level exchange only, and we have also obtained interesting
upper limits on $M_{H^\pm}$ as a function of $\tan\beta$.

Therefore, one can conclude that leptonic $\tau$
decay data constrain 2HDM(II) scenarios with large $\tan\beta$, heavy
$H^\pm$ and light neutral Higgs bosons. 

Obviously, the large 2HDM(II) one-loop corrections found in this paper
can have consequences for other type of processes, which 
will be analysed elsewhere.

\section{Acknowledgement}
The authors thank Theory Group at CERN for a kind hospitality allowing
us to work on a final version of the paper and   M.~J. Herrero for
fruitful discussions. MK is grateful for important 
discussions with M. Misiak, Z. W\c{a}s, B. Stugu and W. Marciano.
This work was partially supported
 by the Polish Committee for Scientific Research,grant  no.~1~P03B~040~26
 and project no.~115/E-343/SPB/DESY/P-03/DWM517/2003-2005
and 
the European Community's Human Potential Programme under contract
HPRN-CT-2000-00149 and and HPRN-CT-2002-00311 EURIDICE.

\end{document}